\documentclass[12pt]{article}
\usepackage{amsmath,amsfonts}
\usepackage{hyperref}
\usepackage{graphicx}
\usepackage{feynmp}
\usepackage{amsthm}
\usepackage{amssymb}
\usepackage{epic}
\usepackage{eepic}
\usepackage[matrix,arrow]{xy}
\usepackage{array}
\usepackage{tikz}
\usetikzlibrary{decorations.pathmorphing}

\usepackage{psfrag}
\usepackage{float}
\usepackage{textcomp}

\makeatletter\@addtoreset{equation}{section}\makeatother

\newcommand{\beq}{\begin{equation}}
\newcommand{\eeq}{\end{equation}}
\newcommand{\bea}{\begin{eqnarray}}
\newcommand{\eea}{\end{eqnarray}}

\renewcommand{\title}[1]{\vbox{\center\LARGE{#1}}\vspace{5mm}}
\renewcommand{\author}[1]{\vbox{\center\large#1}\vspace{5mm}}
\newcommand{\address}[1]{\vbox{\center\em#1}}

\begin{document}
\bibliographystyle{utphys}
\begin{fmffile}{graphs}

\begin{titlepage}
\begin{center}
 
 \vspace{5mm}
 
\hfill {\tt }\\
\vspace{6mm}

\title{\Huge{The AdS/$\mathsf{C}$-$\mathsf{P}$-${\mathsf T}$~Correspondence}}
 
\vspace{7mm}

Jaume Gomis\footnote{\href{mailto:jgomis@perimeterinstitute.ca}{\tt jgomis@perimeterinstitute.ca}} 
 \vskip 10mm
 
\address{
 Perimeter Institute for Theoretical Physics,\\
Waterloo, Ontario, N2L 2Y5, Canada}

\end{center}

\vspace{5mm}
\abstract{
\vspace{2mm}
\noindent
\normalsize{We study   the  realization of     $\mathsf{C}$, $\mathsf{P}$,   $\mathsf{T}$    in ${\cal N}=4$ SYM  -- corresponding to charge conjugation, parity, and time-reversal  --   and identify the    $\mathsf{C}$, $\mathsf{P}$,  $\mathsf{T}$  global symmetries of ${\cal N}=4$ SYM   with        bulk (gauge) symmetries of string theory on  $AdS_5\times S^5$.
The dual   bulk transformations are 
symmetries of Type IIB string theory on $AdS_5\times S^5$ that   combine string worldsheet symmetries,  with geometric transformations acting on $AdS_5\times S^5$. We show that   $\mathsf P$ and $\mathsf T$   map to  $\mathsf{CP}$ and   $\mathsf{CT}$ under $S$-duality (combined with an $SU(4)$ R-symmetry outer automorphism),
while $\mathsf C$ and $ \mathsf{CPT}$ are invariant. We   define     codimension-two  
charge conjugation monodromy defects   in ${\cal N}=4$ SYM, which have an  unusual   large $N$ expansion,  and discuss their bulk   dual   description. We       elucidate     the relation between symmetries   on  the string worldsheet  and symmetries of target space string field theory, global vs gauge symmetries in quantum gravity,  among related   topics. Defects in ${\cal N}=4$ SYM  and in quantum gravity    play a central role in this work.}

\noindent
}
\vfill

\end{titlepage}

\tableofcontents

\section{Introduction}
\label{sec:intro}

The first entry in Maldacena's celebrated AdS/CFT correspondence~\cite{Maldacena:1997re}   (see also  \cite{Gubser:1998bc,Witten:1998qj}) was identifying   the superconformal symmetry algebra of ${\cal N}=4$ super-Yang-Mills (SYM) with the superisometry algebra of Type IIB string  theory on $AdS_5\times S^5$. Shortly thereafter, in \cite{Witten:1998qj}, it was realized that the effective gravitational theory in   $AdS_{5}$   not only contains the $SU(4)$ gauge fields dual to the 
$SU(4)$ R-symmetry of $\mathcal{N}=4$ SYM, but also includes a Chern--Simons term 
whose variation precisely reproduces the  't Hooft anomaly of that R-symmetry. 

In this paper we  answer the following elementary question:  
does $\mathcal{N}=4$ SYM admit $\mathsf{C}$, $\mathsf{P}$ and  $\mathsf{T}$ symmetries and, if so, what is their   dual bulk description?  
$\mathsf{C}$, $\mathsf{P}$ and  $\mathsf{T}$ ($\mathsf{C}$-$\mathsf{P}$-$\mathsf{T}$) transformations -- corresponding to charge conjugation, parity, and time-reversal  -- 
play a fundamental role  as potential discrete global symmetries in relativistic quantum field theory. 
$\mathsf{P}$ and  $\mathsf{T}$ are induced  by the outer automorphisms   of the Lorentz algebra, and 
$\mathsf{C}$, which exchanges particles with antiparticles, correspond, in gauge theories, 
to an outer automorphism of the gauge group.  While the $\mathsf{CPT}$ theorem guarantees the existence of a $\mathsf{CPT}$-symmetry, quantum field theories   -- such as the Standard Model -- need not be invariant under all $\mathsf{C}$-$\mathsf{P}$-$\mathsf{T}$ transformations. 

We briefly summarize here  our findings and refer the reader to section \ref{sec:summary} for extended discussions and outlook.
We show that ${\cal N}=4$ SYM admits well-defined $\mathsf{P}$ and $\mathsf{T}$ symmetry transformations; however, unlike $\mathsf{C}$, these necessarily break the $SU(4)$ R-symmetry.
We demonstrate  that these symmetries behave  nontrivially under $S$-duality:
\beq
\begin{aligned}
  \mathsf{P} &\longleftrightarrow \mathsf{CP}, \\
  \mathsf{T} &\longleftrightarrow \mathsf{CT},
\end{aligned}
\eeq
where $S$ must be accompanied with an outer automorphism of the $SU(4)$ R-symmetry. \newline  $\mathsf P$ and $\mathsf T$ in one duality frame become  $\mathsf{CP}$ and $\mathsf{CT}$ in the S-dual frame,
while $\mathsf C$ and $ \mathsf{CPT}$ are invariant under S-duality.

We  identify the   symmetry transformations of Type IIB string theory on $AdS_5\times S^5$ that are dual  to the $\mathsf{C}$-$\mathsf{P}$-$\mathsf{T}$ transformations of ${\cal N}=4$ SYM. These involve worldsheet symmetries of Type IIB string theory  combined with discrete  spacetime transformations acting on the $AdS_5\times S^5$ geometry. The   AdS/$\mathsf{C}$-$\mathsf{P}$-${\mathsf T}$~Correspondence is:\footnote{We choose the $\mathsf{P}$ outer automorphism of ${\cal N}=4$ SYM to be generated by the reflection $R_3: x^3\rightarrow -x^3$.}
 \beq
 \begin{aligned}
  \mathsf C&\leftrightarrow \Omega(-1)^{F_L}  R_{456789}\,,\\[+2pt]
 \mathsf P&\leftrightarrow (-1)^{F_L}  R_{3456}\,,\\[+2pt]
  \mathsf T&\leftrightarrow  \Omega  R_{0789}\,.
 \label{introdic}
\end{aligned}
 \eeq
$\Omega$ denotes worldsheet parity, $(-1)^{F_L}$  is spacetime fermion parity carried by left-movers on the string worldsheet, and $R_{M}$ is   reflection  along the $x^M=(x^\mu,x^I)$ coordinate of $AdS_5\times S^5$ 
\beq
  ds^2_{AdS_5\times S^5}= X^2 \sum_{\mu=0}^3 \eta_{\mu \nu} dx^\mu dx^\nu+{1\over X^2}\sum_{I=4}^9 dx^I dx^I\,,
  \eeq
  where $X^2=\sum_{I=4}^9 x^I x^I$.  The $\mathsf{CPT}$  symmetry   of ${\cal N}=4$ SYM maps to the 
  $\mathsf{CPT}_{\textrm{bulk}}$  symmetry of quantum gravity on $AdS_5\times S^5$:
 \beq
 \mathsf{CPT}\longleftrightarrow     \mathsf{CPT}_{\textrm{bulk}}\,.
 \label{canoCPT}
 \eeq
 Section \ref{sec:summary} discusses, among other topics,      how the bulk transformations \eqref{introdic} correspond to bulk 
``gauge symmetries", and not global symmetries, as often argued in quantum gravity,  as well as   the relation between worldsheet and target space symmetries in string theory.
We derive the AdS/$\mathsf{C}$-$\mathsf{P}$-${\mathsf T}$~Correspondence  \eqref{introdic} by analyzing the bulk realization of defect operators\footnote{In this paper we consider extended objects  along time (defects) as well as extended objects localized in time (operators). Henceforth, we use interchangeably the word defect and operator.}
 in ${\cal N}=4$ SYM, in terms of  supergravity modes,  branes in the $AdS_5 \times S^5$ background, and asymptotically $AdS_5 \times S^5$ bubbling solutions of Type IIB string theory. The proposed symmetries exactly reproduce the action of $\mathsf{C}$, $\mathsf{P}$, and $\mathsf{T}$ on the bulk description of all defect operators in ${\cal N}=4$ SYM.

 We also discuss how, in any theory with a \(\mathsf{C}\)-symmetry,   codimension-two defects can be defined by requiring that fields undergo a nontrivial charge conjugation monodromy when transported around the defect. This provides an ultraviolet mechanism to define infrared conformal defects since they can be detected from afar. 
We  then  discuss  the bulk $AdS_5\times S^5$ dual description of the charge conjugation    surface defects in ${\cal N}=4$ SYM.
 
\smallskip
The structure of the paper is as follows. In Section~\ref{sec:CPT}, we examine the conditions under which a general $4d$ gauge theory with a charged chiral Weyl fermion, as well as ${\cal N}=4$ SYM, admits $\mathsf{C}$-symmetry. We show that $\mathsf{P}$ and $\mathsf{T}$ symmetries can be realized in ${\cal N}=4$ SYM but necessarily break the $SU(4)$ R-symmetry and analyze the resulting symmetry algebras. We also study the action of $\mathsf{C}$, $\mathsf{P}$, and $\mathsf{T}$ on local, line, and surface defects in ${\cal N}=4$ SYM and determine how these symmetries transform under $S$-duality. In Section~\ref{sec:CPTads}, we investigate the worldsheet and geometric symmetries of Type IIB string theory on $AdS_5 \times S^5$,  and using the bulk description of local, line  and surface operators, we derive   the AdS/$\mathsf{C}$-$\mathsf{P}$-$\mathsf{T}$ correspondence. Section~\ref{sec:surface} explains how charge conjugation symmetry gives rise to codimension-two conformal defects in quantum field theory. We initiate the study of   charge conjugation surface defects in ${\cal N}=4$ SYM, determine some of their basic properties, and discuss their dual description in the $AdS_5 \times S^5$ bulk. Finally, in Section~\ref{sec:summary}, we discuss the broader implications of our results and outline directions for future work, including their relevance to the role of global symmetries in quantum gravity and the interplay between worldsheet and target space symmetries in string theory.

\section{$\mathsf{C}$-$\mathsf{P}$-${\mathsf T}$ Symmetry in ${\cal N}=4$ SYM}
\label{sec:CPT}
 
 The Lagrangian of ${\cal N}=4$  SYM  theory with gauge group $G$  arises from dimensional reduction of $10d$ ${\cal N}=1$ SYM~\cite{Brink:1976bc}:
\begin{equation}
{\cal L} = -\frac{1}{2g^2} \, \text{Tr} \left( F_{MN} F^{MN} + i \lambda^T C\, \Gamma^M D_M \lambda \right),
\label{tendsym}
\end{equation}
supplemented by the topological term
\begin{equation}
\frac{\theta}{8\pi^2} \int \text{Tr} \, F \wedge F,
\label{insta}
\end{equation}
and, when $G$ is not simply connected, by discrete theta angles~\cite{Aharony:2013hda}. 
The gaugino $\lambda$ is a $10d$  Majorana-Weyl spinor. The $\Gamma^M$ matrices satisfy the $\mathrm{Cliff}(1,9)$ Clifford algebra
\begin{equation}
\{\Gamma^M, \Gamma^N\} = 2 \eta^{MN},
\end{equation}
and may be chosen to be real   with $C=\Gamma^0$.  We find it   convenient to formulate ${\cal N}=4$ SYM using  the $10d$ Majorana-Weyl\footnote{\label{susydec} Since ${\bf 16}\rightarrow ({\bf 2},{\bf 1};{ {\bf 4}})\oplus ({\bf 1},{\bf 2};{\bar {\bf 4}})$ under the embedding $\mathrm{Spin}(1,3)\times \mathrm{Spin}(6)\subset \mathrm{Spin}(1,9)$,   $\lambda$ decomposes into  
a pair of $4d$ spinors  $(\lambda_{\alpha}^A, \bar \lambda_{\dot\alpha\, A})$, where  $\alpha ,{\dot\alpha}$ are $\mathrm{Spin}(1,3)$ spinor indices and $A$ is an $\mathrm{Spin}(6)$ spinor index.}  spinor $\lambda$.
 The field content of ${\cal N}=4$ SYM then consists of a gauge field $A_\mu(x)$, a gaugino $\lambda(x)$, and six real scalar fields $\phi^I(x)$, with $\mu = 0,1,2,3$, $I = 4,\dots,9$, and $x \in \mathbb{R}^{1,3}$. All fields transform in the adjoint representation of the gauge group \( G \), which is henceforth taken to be simple, connected, and associated with the Lie algebra \( \mathfrak{g} \).

 ${\cal N}=4$ SYM theory has $PSU(2,2|4)$ superconformal symmetry for 
   any value of the exactly marginal coupling
\beq
\tau={\theta\over 2\pi}+{4\pi   i\over g^2}\,,
\label{tauSYM}
\eeq
where $g^2$ is the gauge coupling of ${\cal N}=4$ SYM.
 The   symmetry is present  for    arbitrary gauge group 
$G$    and  discrete theta angles.     
 The periodicity of $\theta$ depends on the global form of the  gauge group $G$:  it is $2\pi n(G)$ periodic,  where $n(G)$ is an integer that depends on the choice of $G$~\cite{Aharony:2013hda}.

 ${\cal N}=4$ SYM  is endowed with the action of the  duality groupoid $\Gamma\subset SL(2,\mathbb Z)$~\cite{Girardello:1995gf}:\footnote{When  the action on fermionic operators is taken into account, the duality groupoid is a subgroup of   the metaplectic group $Mp(2,\mathbb{Z})$,   a double cover of $SL(2,\mathbb{Z})$~\cite{Pantev:2016nze}. In $\Gamma_0(q)$   $b=0~\text{mod}~q$ while in  $\Gamma^0(q)$ $c=0~\text{mod}~q$.
}
  \beq
 \Gamma:
 \begin{cases}
 SL(2,\mathbb Z)~\qquad \mathfrak g= \mathfrak{su}(N), \mathfrak{so}(2N), \mathfrak{e}_6\\
 PSL(2,\mathbb Z)~~~ ~~\mathfrak g= \mathfrak{su}(2), \mathfrak{e}_7, \mathfrak{e}_8\\
  \Gamma_0(2)~~~~\qquad \mathfrak g= \mathfrak{so}(2N+1), \mathfrak{sp}(N) \\
  \Gamma_0(3)~~~~\qquad \mathfrak g= \mathfrak{g}_2\,,
 \end{cases}
 \eeq
which  acts on $\tau$ by a fractional linear transformation\beq
\tau\rightarrow {a\tau+b\over c\tau+d}\,.
\eeq 
 The action of elements of $\Gamma$ can relate ${\cal N}=4$ SYM theories with different global forms of the gauge group, distinct Lie algebras, and   discrete theta angles~\cite{Gaiotto:2010be,Aharony:2013hda}.\footnote{$\Gamma$ admits a subgroup that, while acting nontrivially on the complexified coupling $\tau$ and the discrete theta angles,    preserves the global form of the gauge group $G$. For $\mathfrak{su}(N)$  this subgroup is $\Gamma_0(N)$ for   $G=SU(N)$ and $\Gamma^0(N)$ for $G=PSU(N)$~\cite{Aharony:2013hda}.}  
 When the simply connected group with Lie algebra $\mathfrak{g}$ 
has a nontrivial center, ${\cal N}=4$ SYM has a one-form symmetry   $\Gamma_{\text{e}} \times \Gamma_{\text{m}}$, which depends on the global form of $G$~\cite{Gaiotto:2014kfa}.

\smallskip
 Does $\mathcal{N}=4$ SYM have additional symmetries for arbitrary value\footnote{Discrete symmetries at special values of $\tau$  have been used to define interesting ${\cal N}=3$ theories in \cite{Garcia-Etxebarria:2015wns,Aharony:2016kai}.} of the gauge coupling $g^2\, $?
Before addressing this question in detail in the remainder of this section by  investigating whether $\mathcal{N}=4$ SYM admits an action of $\mathsf{C}$-$\mathsf{P}$-$\mathsf{T}$ symmetries, we make   some   preparatory remarks.

The $\mathsf{C}\mathsf{P}\mathsf{T}$ theorem guarantees the existence of an (essentially)   canonical $\mathsf{C}\mathsf{P}\mathsf{T}$ symmetry   in any relativistic quantum field theory with a Hamiltonian bounded from below (see \cite{Witten:2025ayw} for a recent lucid account, and     more discussion in section \ref{sec:CPTt}). The $\mathsf{C}\mathsf{P}\mathsf{T}$ transformation is obtained by Wick rotating to the Lorentzian theory on $\mathbb{R}^{1,3}$   a $\mathbb{Z}_2$-transformation  in the connected component of the rotation group acting on the Euclidean geometry $\mathbb{R}^4$. 
 Boundedness of the Hamiltonian is needed for the theory to admit a Wick rotation between Euclidean and Lorentzian signature.  Since any rotation connected to the identity component is, by definition, a symmetry of a relativistic quantum field theory, there is always a $\mathsf{C}\mathsf{P}\mathsf{T}$  symmetry.
 This is  unlike a transformation in the disconnected component of the Lorentz group,  which need not be a symmetry. Indeed,  there is no guarantee that a $\mathsf{C}$, $\mathsf{P}$, or $\mathsf{T}$ symmetry transformation exists in a relativistic quantum field theory.\footnote{These symmetries can emerge as quantum symmetries, see e.g.~\cite{Delmastro:2019vnj}.} For example, the Standard Model does not admit any of these symmetries individually.

  $\mathsf P$ and $\mathsf T$ transformations are induced by the  outer automorphisms of the Poincar\'e algebra acting on $\mathbb R^{1,D-1}$. Their action   on the Poincar\'e generators follows from conjugating the Poincar\'e transformation $x^M\rightarrow a^M+\Lambda^M_{\ N} x^N$ by $\mathsf P$ and $\mathsf T$:
\beq
 \begin{aligned}
 \mathsf P U(a,\Lambda)  \mathsf P^{-1}&=U({\cal P}a, {\cal P}\Lambda {\cal P}^{-1})\,,\\ 
 \mathsf T U(a,\Lambda) \mathsf T^{-1}&=U({\cal T}a, {\cal T}\Lambda{\cal T}^{-1} )\,,\\
 \end{aligned}
 \eeq 
 where $U(a,\Lambda)=e^{ia^M P_M+{i \over 2} w^{MN}M_{MN}}$, and
 \beq
  \begin{aligned}
 {\cal P}x&=(x^0,x^1,\ldots,x^{D-2},-x^{D-1})\,,\\
{\cal T}x&=(-x^0,x^1,\ldots,x^{D-2},x^{D-1})\,.
 \end{aligned}
 \eeq
   Using that $\mathsf P$ is linear and $\mathsf T$ antilinear we obtain the automorphism
 \beq
 \begin{aligned}
  \mathsf P: \begin{cases}
  P_M\rightarrow  ({\cal P}P)_M\\
  M_{MN}\rightarrow ({\cal P}M{\cal P}^{-1})_{MN}
  \end{cases}
  \qquad
   \mathsf T: \begin{cases}
  P_M\rightarrow - ({\cal T}P)_M\\
  M_{MN}\rightarrow -({\cal T}M{\cal T}^{-1})_{MN}\,.
    \end{cases}
  \end{aligned}
 \eeq
 The additional $-$ sign in the action of $\mathsf T$ is due  to its antilinearity.

In a theory possessing $\mathsf{C}$, $\mathsf{P}$, or $\mathsf{T}$ symmetry, these transformations are generally not unique. They can be redefined by composing them with other symmetries of the theory, such as fermion parity, generated by $(-1)^F$, internal flavor symmetries,  Lorentz transformations, $R$-symmetry transformations, etc.  
In  ${\cal N}=4$   SYM $(-1)^F\in \mathrm{Spin}(6)$ R-symmetry as well as belonging to the Lorentz group 
$(-1)^F\in \mathrm{Spin}(1,3)$, giving rise to an $\mathrm{Spin}-SU(4)=(\mathrm{Spin}(1,3)\times SU(4))/\mathbb Z_2$ structure .\footnote{The existence of an $\mathrm{Spin}-SU(4)$ structure allows to consider ${\cal N}=4$ SYM on non-spin manifolds by turning ``twisted flux" obeying  $w_2(TM)=w_2(SO(6))~\textrm{mod}~2$, analogous to in ${\cal N}=2$ theories \cite{Cordova:2018acb,Wang:2018qoy}.}

The discussion of the  $\mathsf{C}$-$\mathsf{P}$-$\mathsf{T}$ 0-form symmetries is most transparent in ${\cal N}=4$ SYM with simply connected gauge group $G=G_{\rm{sc}}$.  In the theory with a non-simply connected  gauge group  $G=G_{\rm{sc}}/K$, obtained by gauging the electric 1-form symmetry group $K\subset \Gamma_{\text{e}}$,  a non-vanishing mixed (0-form)--(1-form) anomaly present in the theory with gauge group $G_{\rm{sc}}$   modifies the nature of the 0-form symmetry in the theory with gauge group $G_{\rm{sc}}/K$.\footnote{Symmetry fractionalization on line operators can affect 0-form anomalies, see e.g.~\cite{Delmastro:2022pfo}.}  The symmetry     becomes noninvertible (see e.g. \cite{Tachikawa:2017gyf,Choi:2021kmx,Kaidi:2021xfk} and review \cite{Schafer-Nameki:2023jdn,Shao:2023gho}).

\subsection{${\mathsf C}$-symmetry}
\label{sec:C}

We start with Yang-Mills theory with gauge group $G$  in arbitrary spacetime dimension.
The transformation 
\beq
{\mathsf C}: A_M^a\rightarrow   O^a_{\, b} A_M^b
\label{ctrans}
\eeq
leaves  the  Yang-Mills Lagrangian  invariant
\beq
{\cal L}_A=-{1\over 2g^2} \text{Tr}F_{MN} F^{MN}=-{1\over 4g^2}\delta_{ab} F^a_{MN} F^{b\,MN}\,,
\label{actionYM}
\eeq
where 
\beq
F^a_{MN}=\partial_M A_{N}^a-\partial_N A_{M}^a+f^a_{bc}A_{M}^bA_{N}^c\,,
\eeq
if the matrix $O$ obeys
\beq
\begin{aligned}
 O^a_{\, d} O^b_{\, e}f_{ab}^c&=O^c_{\, g} f_{de}^g\,,\\[+2pt]
 O^a_{\, c} O^c_{\, b}&=\delta^{a}_b\,.
 \label{automor}
\end{aligned}
\eeq
$f^a_{bc}$ are the structure constants of the Lie algebra $\mathfrak{g}$:
\beq
[T_a,T_b]=i f_{ab}^c T_c\,,
 \eeq
 The   first condition in (\ref{automor})  follows by demanding that $F_{MN}^a$ transforms homogenously under $\mathsf C$, that is    
$F_{MN}^a\rightarrow O^{a}_{\, b}F_{MN}^b$, and the second from imposing invariance of (\ref{actionYM}).

We learn from equation (\ref{automor})   that  Yang-Mills theory is      invariant under the transformation (\ref{ctrans}) if $O^a_{\, b}$ is an orthogonal automorphism of the Lie algebra $\mathfrak g$.  If $O$ is induced by an inner automorphism $\text{Inn}(\mathfrak g)$,  then ${\mathsf C}$  does not generate a  faithfully acting global symmetry, since all  (gauge invariant) operators  are left invariant under the action of ${\mathsf C}$.
   Therefore, the   transformation (\ref{ctrans}) is a global symmetry only   if $O$ is induced by  an outer automorphism of  $\mathfrak g$, which we denote by $\sigma\in \text{Out}(\mathfrak g)$.\footnote{Since $\text{Out}(\mathfrak g)=\text{Aut}(\mathfrak g)/\text{Inn}(\mathfrak g)$  and   $\text{Inn}(\mathfrak g)$ acts trivially on gauge invariant operators, any choice of representative of  $\text{Out}(\mathfrak g)$ acts the same way on gauge invariant operators.}
    The group of outer automorphisms of a simple Lie algebra \( \mathfrak{g} \) is isomorphic to the symmetry group of its Dynkin diagram. 
   The global symmetry generated by ${\mathsf C}$ is then known as charge conjugation.   The action of  ${\mathsf C}$ does not commute with   gauge transformations. 
    
    We now derive the conditions under which Yang-Mills theory coupled to $4d$ fermions possesses charge conjugation symmetry. In $4d$, such a theory can always be formulated in terms of a left-handed Weyl fermion $\psi$ transforming in a (possibly reducible) representation $R$ of $G$. The action is
  \beq
  {\cal L}_\psi=i\psi^\dagger \bar \sigma^\mu D_\mu \psi\,,
  \label{chiral4d}
  \eeq
      where $\bar \sigma^\mu=(1,-\sigma^i)$, with $\sigma^i$ the Pauli matrices. Since  charge conjugation must preserve the chirality of $\psi$, while $\psi^*$ has the opposite chirality,\footnote{This is the reason that any $4d$ theory can be expressed in terms of a left-handed Weyl fermion.} we conclude that the most general charge conjugation transformation acting on a Weyl fermion is
\begin{equation}
{\mathsf C}: \psi \rightarrow U B \psi \,,
\label{ctransf}
\end{equation}
where $U$ is a unitary matrix acting on the representation indices of $\psi$ and $B$ acts on its spinor indices.  Invariance of the kinetic term    requires that $B^\dagger \bar\sigma^\mu B= \bar\sigma^\mu$, and Schur's lemma implies that $B\propto \mathbb{I}$.  Combining  \eqref{ctransf} with  the gauge field transformation (\ref{ctrans}), the      fermion action \eqref{chiral4d} has charge conjugation symmetry if 
  \beq
   U^\dagger O^a_{\, b} T^b_RU=T^a_R\,.
   \label{condf}
   \eeq 
  $T^a_R$ are the generators of $\mathfrak{g}$ in the representation $R$, while
\beq
O^a_{\, b} T^b_R=T^a_{\sigma(R)}
\eeq
are the generators  of $\mathfrak{g}$ in the conjugate representation $\sigma(R)$. The Dynkin labels of the representation $\sigma(R)$ are obtained by permuting those of $R$ according to the symmetry of the Dynkin diagram generated by $\sigma$. 
  Equation (\ref{condf}) says that Yang-Mills theory  coupled to a $4d$ Weyl fermion $\psi$ has ${\mathsf C}$-symmetry if $R$ is self-conjugate under the action of the outer automorphism.
 The condition for ${\mathsf C}$-symmetry is therefore    that
\beq
\sigma(R)= R\, 
\label{condrep}
\eeq
up to unitary equivalence. This implies, contrary to   statements often found in the literature, that $4d$ chiral gauge theories {\it can}   have charge conjugation symmetry. 

A very similar analysis can be carried out for a scalar field   in a representation $R$, with action
 \beq
 {\cal L}_\phi= (D_\mu \phi)^\dagger D^\mu \phi\,.
 \eeq
Invariance under charge conjugation requires $R$ to be self-conjugate, that is it must obey  \eqref{condrep}. Yukawa couplings can be constrained by    \( \mathsf{C} \)-symmetry.  In ${\cal N}=4$ SYM, the Yukawa coupling  $\hbox{Tr}\left(\lambda^T \Gamma^0 \Gamma^I[\phi^I,\lambda]\right)$  fixes the charge conjugation parity of the scalar fields $\phi^I$.

Since the ${\cal N}=4$ SYM     fields are in the adjoint representation, and   the adjoint representation is self-conjugate under any automorphism $\sigma$, the gaugino $\lambda$  and scalars $\phi^I$  automatically obey  (\ref{condrep}). Therefore,  ${\cal N}=4$ SYM with gauge group $G$ has a charge conjugation symmetry $\mathsf C$ if and only if the Lie algebra $\mathfrak{g}$ of $G$ admits an outer automorphism. 
Consequently, ${\cal N}=4$ SYM
 based on the Lie algebra    $\mathfrak{su}(N)$ for $N\geq 3$, $\mathfrak{so}(2n)$ and $\mathfrak e_6$ has a $\mathbb Z^{\mathsf C}_2$ charge conjugation symmetry, and no ${\mathsf C}$-symmetry for other gauge groups.\footnote{For $\mathfrak{so}(8)$ it is  an enhanced to an $S_3$ symmetry.}

The \( \mathbb{Z}_2^{\mathsf{C}} \) global symmetry generated by \( \mathsf{C} \) can be taken to act on the \( \mathcal{N}=4 \) SYM fields as follows:
  \beq
\begin{aligned}
\mathsf{C}:
\begin{cases}
 \mathfrak{su}(N): \Upsilon\rightarrow -\Upsilon^T\qquad N>2\\[+2pt]
 \mathfrak{so}(2N):\Upsilon\rightarrow {P}\Upsilon{P}^T\\[+2pt]
  \mathfrak{e}_6: ~~~~~~\Upsilon\rightarrow -\Upsilon^T\,,
  \end{cases}
\end{aligned}
\label{transfo}
\eeq
where we have assembled the ${\cal N}=4$ SYM fields into a 
  vector 
\beq
\Upsilon=(A_\mu, \lambda, \phi^I)\,.
\label{collective}
\eeq 
  \( T \) denotes matrix transposition, and \( {P} = \mathrm{diag}(-1, 1, \ldots, 1) \) is an improper element of \( O(2N) \). This a symmetry of ${\cal N}=4$ SYM for arbitrary complexified coupling $\tau$, which we express as 
    \begin{equation}
\mathsf{C}:   \tau \rightarrow  \tau \,,
\label{Contau}
\end{equation}
   and    for arbitrary discrete theta angles. 
The  transformation (\ref{transfo}) leaves the $\mathcal{N}=4$  SYM  $PSU(2,2|4)$ Noether currents invariant, and generates  a $\mathbb Z^{\mathsf C}_2$ global symmetry.\footnote{The transformation \eqref{transfo} can be combined with the action of $\mathbb Z_2\subset SU(4)$ $R$-symmetry  or equivalenty $(-1)^F$. Our choice is the unique one preserving $PSU(2,2|4)$. }   
 
 We  turn now to the action of the charge conjugation symmetry \( \mathbb{Z}_2^{\mathsf{C}} \) on local, line, and surface operators in \( \mathcal{N}=4 \) SYM, focusing on  \( \mathfrak{g} = \mathfrak{su}(N) \).

 \medskip
 \noindent
 $\bullet$ Local operators
  \medskip
  
Consider the short multiplets of \( \mathcal{N}=4 \) SYM. The bottom component of each such multiplet is a scalar chiral primary operator \( \mathcal{O}_\Delta \) of scaling dimension \( \Delta \), transforming in the \( (0, \Delta, 0) \) representation of the \( SU(4) \) R-symmetry. The   values of \( \Delta \) are  the degrees of the Casimir invariants of the   Lie algebra \( \mathfrak{g} \).  Since \( \mathbb{Z}_2^{\mathsf{C}} \) commutes with the \( PSU(2,2|4) \) superconformal algebra, all operators within a given multiplet transform under \( \mathsf{C} \) in the same way as the bottom component \( \mathcal{O}_\Delta \).

The chiral primary operators for  \(\mathfrak{su}(N) \) are
 \beq
 \mathcal{O}_\Delta = C^\Delta_{I_1\ldots I_\Delta}\text{Tr}\left(\phi^{I_1}\ldots \phi^{I_\Delta}\right)  \qquad\Delta=2,3,\ldots, N\,,
 \eeq
 where $C^\Delta_{I_1\ldots I_\Delta}$ is a totally symmetric tensor. By virtue of the symmetry of $C^\Delta_{I_1\ldots I_\Delta}$, $\mathcal{O}_\Delta$ is diagonal under the action of $\mathsf{C}$ in (\ref{transfo})
 \beq
{ \mathsf C}\left(\mathcal{O}_\Delta\right)=(-1)^\Delta \mathcal{O}_\Delta\,.
 \label{actonCPO}
 \eeq
Operators with $\Delta=\text{odd}$ are charged under $\mathsf{C}$, while those with $\Delta=\text{even}$ are neutral.

 \medskip
 \noindent
 $\bullet$ Line operators
 \medskip
 
\( \mathcal{N}=4 \) SYM with gauge group $G$ has supersymmetric Wilson, 't Hooft   and   Wilson-'t Hooft line operators  labeled by  a pair of weights~\cite{Kapustin:2005py}
\beq
(\mu,\nu)\in (\Lambda_{\text{char}}\times\Lambda_{\text{cochar}})/{\cal W} \, 
\eeq
obeying the mutual locality condition
\beq
\langle \mu,\nu \rangle\in \mathbb Z\,,
\eeq
where \( \langle \cdot, \cdot \rangle \) denotes the canonical pairing between the character and cocharacter lattices, \( \Lambda_{\text{char}} \) and \( \Lambda_{\text{cochar}} \). ${\cal W}$ is the Weyl group of $G$.
 
 To analyze the action of \(\mathbb{Z}_2^{\mathsf{C}}\) on line operators, it is useful to begin by considering the maximal lattice of electric and magnetic charges.  For electric charges, this corresponds to  taking the simply connected form of the gauge group corresponding to $\mathfrak g$.
 The magnetic charges are maximized by taking the simply connected form of the gauge group corresponding to the Langlands (or GNO) dual  Lie algebra ${}^L\mathfrak g$.
  A particular choice of the global form of \( G \) and of discrete theta angles then selects a maximal, mutually local subset of these charges, which defines an allowed spectrum of Wilson and 't Hooft line operators~\cite{Gaiotto:2010be,Aharony:2013hda}.

\smallskip
 
\noindent
\textit{Wilson lines:} Consider a Wilson line operator labeled by a representation \( R \) of the simply connected form of the gauge group corresponding to $\mathfrak g$. It is given by~\cite{Maldacena:1998im,Rey:1998ik}
\begin{equation}
W_R = \operatorname{Tr}_R \, \mathcal{P} \exp  \left(i \int    A_\mu \, dx^\mu + \phi \, |\dot{x}| \, d\tau  \right)\,.
\label{Wilsonf}
\end{equation}
The Wilson loop couples to a   scalar field \( \phi \), chosen arbitrarily from the six scalars \( \phi^I \) of the theory.
 
Under the action of an outer automorphism \( \sigma \), Wilson lines transform as
\begin{equation}
W_R \rightarrow W_{\sigma(R)}\,,
\end{equation}
with \( \sigma(R) \)   the representation obtained by acting on \( R \) with \( \sigma \). 
Wilson loops in self-conjugate representations are invariant under \( \mathsf{C} \), while those in non-self-conjugate representations form a     pair exchanged by \( \mathsf{C} \).

For $\mathfrak{g}=\mathfrak{su}(N)$ 
 \begin{equation}
{\mathsf C}(W_R) = W_{ \overline R}\,,
\label{CwilsonR}
\end{equation}
where \( \overline{R} \) denotes the complex conjugate representation, characterized by the property that \( R \otimes \overline{R} \) contains the trivial representation. For future reference, we  record how \( \mathsf{C} \) acts on the fundamental representation $\text{fund}$, the rank $k$  antisymmetric $\text{A}_k$ and the rank $k$ symmetric $\text{S}_k$ representation  Wilson lines:
\beq
\begin{aligned}
{\mathsf C}(W_{\text{fund}}) &= W_{ \overline {\text{fund}}}\,,\\
{\mathsf C}(W_{\text{A}_k}) &= W_{   {\text{A}_{N-k}}}\,,\\
{\mathsf C}(W_{\text{S}_k}) &=   W_{\overline{\mathrm{S}_{k}}}\,.
\label{Cwilson}
\end{aligned}
\eeq
 \smallskip

\noindent
\textit{'t Hooft lines:} Consider a 't Hooft line defetc along $x^0$ labeled by a highest weight $B$ of a representation \(R \) of the simply connected form of the gauge group with Langlands dual Lie algebra
${}^{L}{\mathfrak{g}} \). It is defined by the following boundary condition near the line defect~\cite{Kapustin:2005py} 
\begin{equation}
  F_{ij}(x)={B\over 2}{\epsilon_{ijk}x^k\over |x|^3}\,,\qquad~~ \phi(x)={B\over 2 |x|} {g^2\over 4\pi}|\tau|\,.
  \label{thooftsingu}
\end{equation}
The 't Hooft loop couples to a   scalar field \( \phi \), also chosen arbitrarily from the six scalars \( \phi^I \). 
 
 Under the action of the outer automorphism \( \sigma \) defined in (\ref{transfo}), the 't Hooft line operators transform as
\begin{equation}
T_R \rightarrow T_{\sigma(R)}\,. 
\end{equation}
 For $\mathfrak{g}=\mathfrak{su}(N)$ 
 \begin{equation}
{\mathsf C}(T_R) = T_{ \overline R}\,.
\label{conTR}
\end{equation}
 As before, we  record how \( \mathsf{C} \) acts on the fundamental, antisymmetric and symmetric representation 't Hooft lines
\beq
\begin{aligned}
{\mathsf C}(T_{\text{fund}}) &= T_{ \overline {\text{fund}}}\,,\\
{\mathsf C}(T_{\text{A}_k}) &= T_{   {\text{A}_{N-k}}}\,,\\
{\mathsf C}(T_{\text{S}_k}) &= T_{  \overline { {\text{S}_{k}}}}\,.
\end{aligned}
\label{conT}
\eeq
  
 \medskip
 \noindent
 $\bullet$ Surface  operators
 \medskip
 
\( \mathcal{N}=4 \) SYM with gauge group $G$ has supersymmetric  surface  operators ${\cal O}_{\Sigma}(\alpha,\beta,\gamma,\eta)$ labeled by a choice of Levi subgroup $\mathbb L\subset G$ and a collection of continuous parameters 
\beq
(\alpha,\beta,\gamma,\eta)\in (\mathbb T\times \mathfrak t \times \mathfrak t\times {}^{L}\mathbb T)/{\cal W}_{\mathbb L}\,,
\label{modulis}
\eeq
where $\mathbb T$ and ${}^{L}\mathbb T$ are the maximal tori of $G$ and Langlands dual group ${}^LG$,  $\mathfrak{t}$ is the Cartan subalgebra of $\mathfrak{g}$ and ${\cal W}_{\mathbb L}$ the Weyl group of $\mathbb L$~\cite{Gukov:2006jk}.  
${\cal O}_{\Sigma}(\alpha,\beta,\gamma,\eta)$ supported the surface  $\Sigma=\mathbb R^{1,1}\subset \mathbb R^{1,3}$ is defined by the following boundary condition as the operator is approached~\cite{Gukov:2006jk}
\beq
A=\alpha d\theta\,,\qquad \Phi={\beta+i\gamma\over z}\,,
\label{surface1}
\eeq
together with the insertion  in the functional integral of 
\beq
\exp\left(i\int_{\Sigma} \text{Tr}\, \eta F  \right)\,,
\label{surface2}
\eeq
where $z=|z|e^{i\theta}$ is the coordinate transverse to the defect. The surface defect excites a complex scalar $\Phi$ chosen arbitrarily from the six scalars \( \phi^I \).  

 Under the action of the outer automorphism \( \sigma \) defined in (\ref{transfo}),  surface operators in  $  \mathfrak{su}(N)$ ${\cal N}=4$ SYM transform into each other as follows:
 \beq
 \mathsf C\left({\cal O}_{\Sigma}(\alpha,\beta,\gamma,\eta)  \right)={\cal O}_{\Sigma}(-\alpha,-\beta,-\gamma,-\eta)\,.
 \label{actsurf}
 \eeq
 Since  $-1\notin {\cal W}_{\mathbb L}$ for $\mathfrak{g}=  \mathfrak{su}(N)$, the action of $\mathsf C$ on surface operators is nontrivial (cf. \eqref{modulis}).

\subsection{${\mathsf P}$-symmetry}
\label{sec:P}

Our discussion begins with the observation that $10d$ ${\cal N}=1$ SYM is not parity invariant.  Indeed, reversing an odd number of spatial coordinates in $\mathbb{R}^{1,9}$ flips the chirality of the $10d$ Majorana-Weyl fermion $\lambda$, which   does not preserve the field content of the theory. As a consequence, the theory is also not invariant under the combined transformation $\mathsf{CP}$, and hence not invariant under $\mathsf{T}$ either, by the $\mathsf{CPT}$ theorem.

 This obstruction actually suggests an   strategy for defining $\mathsf{P}$-symmetry in ${\cal N}=4$ SYM. The theory arises via dimensional reduction of $10d$ ${\cal N}=1$ SYM, where spacetime splits as $\mathbb{R}^{1,9} \to \mathbb{R}^{1,3} \times \mathbb{R}^6$. We can define parity in ${\cal N}=4$ SYM   as a transformation that simultaneously reverses an  odd number of coordinates  in both $\mathbb{R}^{1,3}$ and $\mathbb{R}^6$. This combined reflection preserves the chirality of the $10d$ spinor $\lambda$. Crucially, while such a transformation lies in the  connected component  of the Lorentz group acting on $\mathbb{R}^{1,9}$, and is therefore a symmetry, its dimensional reduction does \emph{not} belong to the connected component of the Lorentz group acting on  $\mathbb{R}^{1,3}$. This   
  makes it apparent that there is no way to define  $\mathsf{P}$-symmetry in ${\cal N}=4$ SYM while preserving the $Spin(6)$ R-symmetry of ${\cal N}=4$ SYM.  
  
 We define the action of parity   on the  ${\cal N}=4$ SYM fields as
 \beq
 \mathsf P:
 \begin{cases}
 A_i(x)\rightarrow A_i({\cal P} x)\qquad ~~~i=0,1,2\\
  A_3(x)\rightarrow -A_3({\cal P} x)\\
 \phi^A(x)\rightarrow -\phi^A({\cal P} x)\qquad A=4,5,6\\
 \phi^{\dot A}(x)\rightarrow  \phi^{\dot A}({\cal P} x)\qquad ~~\dot A=7,8,9\\
\lambda(x)\rightarrow\Gamma^{3456}\, \lambda({\cal P} x)\,,
 \end{cases}
 \label{p1act}
 \eeq
  where 
  \beq
  {\cal P} x\equiv (x^0,x^1,x^2,-x^3)\, 
 \eeq
 is a reflection of the $x^3$ coordinate in $\mathbb R^{1,3}$.
 The six scalars $\phi^I$ decompose under $Spin(6)\to Spin(3)\times Spin(3)$ as $\phi^I = (\phi^A, \phi^{\dot A})$, with each triplet transforming as a vector under one of the $Spin(3)$ factors.  $\phi^A$ are parity pseudo-scalars while $\phi^{\dot A}$ are scalars.  

It is straightforward to verify that the dimensionally reduced Lagrangian~(\ref{tendsym}) is invariant under the transformation~(\ref{p1act}) by virtue of $B=\Gamma^{3456}$ obeying
\beq
 \begin{aligned}
\Gamma^0 B^T \Gamma^iB&=\Gamma^0\Gamma^i\,,\\
\Gamma^0 B^T\Gamma^3\Gamma^iB&=-\Gamma^0\Gamma^3\,,\\
  B^T\Gamma^{0A} B&= -\Gamma^{0A}\,,\\
 B^T\Gamma^{0\dot A} B&=  \Gamma^{0\dot A} \,.
 \end{aligned}
 \eeq
 We use  that  $\Gamma^0$ is antisymmetric while the rest are symmetric.
\eqref{p1act} thus  defines
   a $\mathbb Z^{\mathsf P}_2$ global symmetry for arbitrary Yang-Mills coupling $g^2$.

   The symmetry transformation~(\ref{p1act})  obeys $\mathsf{P}^2=1$. 
This implies that ${\cal N}=4$ SYM can be consistently formulated on an  unorientable  manifold   admitting a $\mathrm{Pin}^+$ structure~\cite{Kapustin:2014dxa,Witten:2015aba}. Any other choice of parity  is obtained  from (\ref{p1act}) by combining it with a transformation belonging to the connected component of the Lorentz group acting on 
 $\mathbb R^{1,3}\times \mathbb R^{6}$. Doing this, it is possible\footnote{For example, by composing \eqref{p1act} with a   $\mathrm{Spin}(6)$ transformation, we can define a   parity symmetry $\widetilde{\mathsf{P}}$ such that $\widetilde{\mathsf{P}}: \lambda(x) \rightarrow \Gamma^{345678} \lambda({\cal P}x)$.} to define a parity symmetry transformation $\widetilde{\mathsf{P}}$ with 
 \beq
 \widetilde{\mathsf{P}}^2 = (-1)^F\,,
 \eeq
 where $(-1)^F$ is $4d$ fermion parity. 
  With this version of parity, ${\cal N}=4$ SYM can instead be defined on an unoriented  manifold   admitting a $\mathrm{Pin}^-$ structure~\cite{Kapustin:2014dxa,Witten:2015aba}.

In contrast, the topological term~(\ref{insta}) changes sign under this transformation. This can be equivalently expressed by saying that   $\mathsf{P}$ acts on the complexified coupling $\tau$ as follows:
\begin{equation}
\mathsf{P}:   \tau \rightarrow -\bar{\tau}\,.
\label{Pontau}
\end{equation}
 ${\cal N}=4$ SYM with gauge group $G$ is therefore parity invariant   for $\theta=0$ or $\theta= {n(G)}\pi$, where $2\pi n(G)$ is the periodicity of $\theta$ in the theory with gauge group $G$.  For these values of \( \theta \), the topological term~(\ref{insta}) contributes a factor of \( +1 \) or \( -1 \) in the sum over gauge field configurations. Similarly, \( \mathsf{P} \)-symmetry imposes constraints on discrete theta angles. These angles contribute phases to the functional integral when summing over nontrivial topological sectors of \( G \)-bundles that cannot be lifted to bundles of the simply connected group associated with the Lie algebra \( \mathfrak{g} \). Only those discrete theta angles whose contributions are   \( \pm 1 \) are compatible with \( \mathsf{P} \)-symmetry. This can be easily understood when the theory is Wick rotated to Euclidean signature. All the topological terms are purely imaginary and under a change of orientation they flip sign. There is   a bonafide symmetry only when they produce contributions that are purely real, that is \( \pm 1 \).

 We  turn now to the action of the  parity symmetry \( \mathbb{Z}_2^{\mathsf{P}} \) on local, line, and surface operators in \( \mathcal{N}=4 \) SYM, focusing on  \( \mathfrak{g} = \mathfrak{su}(N) \).

 \medskip
 \noindent
 $\bullet$ Local operators
  \medskip

  The action of parity~\eqref{p1act} on the chiral primary operators of \( \mathcal{N}=4 \) SYM can be obtained by decomposing the \( (0, \Delta, 0) \) representation of \( \mathrm{Spin}(6) \) into representations of \( \mathrm{Spin}(3) \times \mathrm{Spin}(3) \). Each operator in this decomposition transforms in a spin \((j_1, j_2)\) representation under\footnote{This follows from the branching rule \( \mathbf{6} \to (\mathbf{1}, \mathbf{0}) \oplus (\mathbf{0}, \mathbf{1}) \).} \( \mathrm{Spin}(3) \times \mathrm{Spin}(3) \), with \( j_1, j_2 \in \mathbb{Z} \). Denoting such an operator by \( \mathcal{O}_\Delta^{j_1, j_2} \),       parity acts as
\begin{equation}
\mathsf{P} \left( \mathcal{O}_\Delta^{j_1, j_2}(x) \right) = (-1)^{j_1} \, \mathcal{O}_\Delta^{j_1, j_2}(\mathcal{P} x) \,.
\label{parityCPO}
\end{equation}
There are $\mathsf{P}$-odd operators for every value of $\Delta$.

 \medskip
 \noindent
 $\bullet$ Line operators
 \medskip
 
There are two qualitatively distinct types of line operators, distinguished by how they are embedded in \( \mathbb{R}^{1,3} \). Without loss of generality, the operator can lie along the time direction\footnote{Any line operator lying in the fixed locus \( \mathbb{R}^{1,2} \) at \( x^3 = 0 \) behaves identically under \( \mathsf{P} \).} \( x^0 \), or along the spatial direction \( x^3 \), which is inverted under parity. We distinguish the latter case by placing a hat on the operator.

 We focus on Wilson and 't Hooft line operators that transform simply under parity. These couple  to an  appropriate scalar field  with either \( \phi \in \phi^A \) or \( \phi \in \phi^{\dot A} \). We label a line operator by the scalar field type it couples to.

\smallskip
 
\noindent
\textit{Wilson lines:}  The transformation of a Wilson line operator~\eqref{Wilsonf} under \( \mathsf{P} \) is given by:
\begin{equation}
\begin{aligned}
\mathsf{P}(W_{R}(\phi^{\dot A})) &= W_{R}(\phi^{\dot A}) \,, \\
\mathsf{P}(\hat W_{R}(\phi^{A})) &= \hat W_{\overline{R}}(\phi^{A}) \,,
\end{aligned}
\end{equation}
while its transformation under \( \mathsf{C}\mathsf{P} \) is:
\begin{equation}
\begin{aligned}
\mathsf{C}\mathsf{P}(W_{R}(\phi^{\dot A})) &= W_{\overline{R}}(\phi^{\dot A}) \,, \\
\mathsf{C}\mathsf{P}(\hat W_{R}(\phi^{A})) &= \hat W_{R}(\phi^{A}) \,.
\end{aligned}
\end{equation}

\smallskip
 
\noindent
\textit{'t Hooft lines:} 

The parity properties of 't Hooft line operators are determined by analyzing how the singularity in equation~\eqref{thooftsingu} transforms under the action of \( \mathsf{P} \). For a line operator extending along the \( x^0 \) direction and coupled to a scalar \( \phi \in \phi^{A} \), the parity transformation acts as
\begin{equation}
\mathsf{P}:
\begin{cases}
F_{12}(x) \rightarrow F_{12}(\mathcal{P}x) = -F_{12}(x) \,, \\
F_{31}(x) \rightarrow -F_{31}(\mathcal{P}x) = -F_{31}(x) \,, \\
F_{23}(x) \rightarrow -F_{23}(\mathcal{P}x) = -F_{23}(x) \,, \\
\phi(x) \rightarrow -\phi(\mathcal{P}x) = -\phi(x) \,.
\end{cases}
\end{equation}
In the final equality of each line, we have used the explicit form of the singularity given in equation~\eqref{thooftsingu}.
For a line operator along $x^3$ and with $\phi\in \phi^{\dot A}$
\beq
 {\mathsf P}:
  \begin{cases}
F_{ij}(x)\rightarrow F_{ij}({\cal P}x)=F_{ij}(x)\qquad i,j=0,1,2\\
\phi(x)\rightarrow \phi({\cal P}x)= \phi(x)\,.
\end{cases}
\eeq
 Therefore, the transformation of 't Hooft line operators under \( \mathsf{P} \) is given by:
\begin{equation}
\begin{aligned}
\mathsf{P}(T_{R}(\phi^{A})) &= T_{\overline{R}}(\phi^{A}) \,, \\
\mathsf{P}(\hat{T}_{R}(\phi^{\dot{A}})) &= \hat{T}_{R}(\phi^{\dot{A}}) \,,
\end{aligned}
\end{equation}
while under \( \mathsf{C}\mathsf{P} \), we have:
\begin{equation}
\begin{aligned}
\mathsf{C}\mathsf{P}(T_{R}(\phi^{A})) &= T_{R}(\phi^{A}) \,, \\
\mathsf{C}\mathsf{P}(\hat{T}_{R}(\phi^{\dot{A}})) &= \hat{T}_{\overline{R}}(\phi^{\dot{A}}) \,.
\end{aligned}
\end{equation}

The results summarized in Table~\ref{linePCP} can be used to determine how the discrete symmetries $\mathsf{P}$ and $\mathsf{CP}$ behave under S-duality. 
The $S$-transformation in 
$\Gamma\subset SL(2,\mathbb{Z})$ maps a Wilson line in a representation \( R \) of a gauge group \( G \) to a 't~Hooft line labeled by a magnetic weight of the Langlands dual group \( {}^L G \)~\cite{Kapustin:2005py}.\footnote{Chiral primary operators, suitably normalized, are singlets under $S$-duality.} This magnetic weight corresponds to the highest weight of the original representation \( R \). In short:
\beq
S: \quad W_R \;\longrightarrow\; T_R\,.
\eeq
Therefore, the action of $S$,   combined with the outer automorphism
\beq
  \mathcal{R} : \mathrm{Spin}(3) \longleftrightarrow \mathrm{Spin}(3)
  \label{outerSU4}
\eeq
exchanging the two $\mathrm{Spin}(3)$ subgroups of $\mathrm{Spin}(6)\supset \mathrm{Spin}(3)\times \mathrm{Spin}(3)$, swaps the two symmetries:
\beq
  S\,\mathcal{R} : \quad \mathsf{P} \longleftrightarrow \mathsf{CP}\,,
  \label{CtoCP}
\eeq
or more  explicitly:
\beq
(S\,\mathcal{R}) \, \mathsf{P} \, (S\,\mathcal{R} )^{-1}=\mathsf{CP}\,.
  \label{CtoCPex}
\eeq
  \begin{table}[h!]
\centering
\renewcommand{\arraystretch}{1.4}
\begin{tabular}{|c|c|c|}
\hline
\text{Line Operator} & \textbf{  \( \mathsf{P} \)} & \textbf{  \( \mathsf{CP} \)} \\
\hline
\( W_{R}(\phi^{\dot{A}}) \) & \( W_{R}(\phi^{\dot{A}}) \) & \( W_{\overline{R}}(\phi^{\dot{A}}) \) \\
\hline
\( T_{R}(\phi^{A}) \) & \( T_{\overline{R}}(\phi^{A}) \) & \( T_{R}(\phi^{A}) \) \\
\hline
\( \hat{W}_{R}(\phi^{A}) \) & \( \hat{W}_{\overline{R}}(\phi^{A}) \) & \( \hat{W}_{R}(\phi^{A}) \) \\
\hline
\( \hat{T}_{R}(\phi^{\dot{A}}) \) & \( \hat{T}_{R}(\phi^{\dot{A}}) \) & \( \hat{T}_{\overline{R}}(\phi^{\dot{A}}) \) \\
\hline
\end{tabular}
\caption{Transformation of Wilson and 't Hooft lines under \( \mathsf{P} \) and \( \mathsf{CP} \).}
\label{linePCP}
\end{table}

This is in contrast to charge conjugation, which is invariant under $S$-duality:
\beq
(S\,\mathcal{R}) \, \mathsf{C} \, (S\,\mathcal{R} )^{-1}=\mathsf{C}\,,
  \label{CtoC}
\eeq
as can be inferred from the results of section \ref{sec:C}.

 \medskip
 \noindent
 $\bullet$ Surface  operators
 \medskip

Surface operators supported on the two-dimensional plane $\Sigma = \mathbb{R}^{1,1}$, with local coordinates $(x^0, x^3)$, admit a simple action under the parity symmetry $\mathsf{P}$. Each operator couples to a complex scalar field as per \eqref{surface1}, where one chooses either
\beq
  \Phi \in \phi^A
  \quad\text{or}\quad
  \Phi \in \phi^{\dot A}.
\eeq
We attach to each surface operator a label indicating which complex scalar it couples to.

The parity properties of   surface operators follow from how the singularity in \eqref{surface1} and the defect insertion in \eqref{surface2} transform under the action of \(\mathsf{P}\) in \eqref{p1act}.  Explicitly, we find  that under \(\mathsf{P}\):\footnote{Under $\mathsf P$, the electric field $F_{03}$ along $\mathbb R^{1,1}$ is odd while  the magnetic field $F_{12}$ is even.}
\begin{equation}
\begin{aligned}
\mathsf{P}\bigl(\mathcal{O}_\Sigma(\alpha,\beta,\gamma,\eta)[\phi^{\dot A}]\bigr)
  &= \mathcal{O}_\Sigma(\alpha,\beta,\gamma,-\eta)[\phi^{\dot A}], \\[6pt]
\mathsf{P}\bigl(\mathcal{O}_\Sigma(\alpha,\beta,\gamma,\eta)[\phi^{A}]\bigr)
  &= \mathcal{O}_\Sigma(\alpha,-\beta,-\gamma,-\eta)[\phi^{A}],
\end{aligned}
\label{actonsurfpp}
\end{equation}
and under \(\mathsf{CP}\):
\begin{equation}
\begin{aligned}
\mathsf{CP}\bigl(\mathcal{O}_\Sigma(\alpha,\beta,\gamma,\eta)[\phi^{\dot A}]\bigr)
  &= \mathcal{O}_\Sigma(-\alpha,-\beta,-\gamma,\eta)[\phi^{\dot A}], \\[6pt]
\mathsf{CP}\bigl(\mathcal{O}_\Sigma(\alpha,\beta,\gamma,\eta)[\phi^{A}]\bigr)
  &= \mathcal{O}_\Sigma(-\alpha,\beta,\gamma,\eta)[\phi^{A}].
\end{aligned}
\end{equation}

 This allows us to understand how $\mathsf{P}$ and $\mathsf{CP}$ behave under S-duality. The element $S$  in $\Gamma\subset SL(2,\mathbb{Z})$ acts on the  surface operator  parameters as~\cite{Gukov:2006jk,Gomis:2007fi}
\beq
  S:
  \begin{cases}
    (\alpha,\eta)\;\longmapsto\;(\eta,\,-\alpha)\,,\\[6pt]
    (\beta,\gamma)\;\longmapsto\;|\tau|\;(\beta,\gamma)\,.
  \end{cases}
\label{actonsurf}
\eeq
This implies, using the  results summarized in Table~\ref{surfacePCP}, that indeed $\mathsf{P}$ and $\mathsf{CP}$ are 
exchanged by $S$-duality as in~\eqref{CtoCP} and  that:
\beq
(S\,\mathcal{R})\,  \mathsf{P}\,  (S\,\mathcal{R} )^{-1}=\mathsf{CP}\,.
\eeq

\begin{table}[ht]
\centering
\renewcommand{\arraystretch}{1.4}
\begin{tabular}{|c|c|c|}
\hline
Surface Operator 
  & $\mathsf{P}$ 
  & $\mathsf{CP}$ \\ 
\hline
$\displaystyle \mathcal{O}_\Sigma(\alpha,\beta,\gamma,\eta)[\phi^{\dot A}]$
  & 
  $\displaystyle \mathcal{O}_\Sigma(\alpha,\beta,\gamma,-\eta)[\phi^{\dot A}]$
  &
  $\displaystyle \mathcal{O}_\Sigma(-\alpha,-\beta,-\gamma,\eta)[\phi^{\dot A}]$
  \\[2ex]
\hline
$\displaystyle \mathcal{O}_\Sigma(\alpha,\beta,\gamma,\eta)[\phi^{A}]$
  &
  $\displaystyle \mathcal{O}_\Sigma(\alpha,-\beta,-\gamma,-\eta)[\phi^{A}]$
  &
  $\displaystyle \mathcal{O}_\Sigma(-\alpha,\beta,\gamma,\eta)[\phi^{A}]$
  \\ 
\hline
\end{tabular}
\caption{Transformation  of surface operators under   $\mathsf{P}$ and $\mathsf{C}\mathsf{P}$.}
\label{surfacePCP}
\end{table}

\subsection{${\mathsf T}$-symmetry}
\label{sec:T}

Although $10d$ ${\cal N}=1$ SYM is not invariant under time-reversal ${\mathsf T}$, this non-invariance can be remedied upon dimensional reduction to $4d$ ${\cal N}=4$ SYM by combining time-reversal in $\mathbb{R}^{1,3}$ with a suitable parity transformation acting on the internal space $\mathbb{R}^{6}$. A necessary condition on the parity transformation in $\mathbb{R}^{6}$ is that it must reverse an odd number of coordinates, ensuring that the combined time-reversal transformation $\mathsf{T}$ preserves the chirality of the gaugino $\lambda$.
Any such  transformation lies in the  connected component   $\mathrm{Spin}(1,9)$   but not in 
the connected component of $\mathrm{Spin}(1,3)$, and thus defines an action of ${\mathsf T}$ in $4d$.

 We implement the action of time-reversal   on the  ${\cal N}=4$ SYM fields as
 \beq
 \mathsf T:
 \begin{cases}
 A_0(x)\rightarrow A_0({\cal T} x)\qquad \\
  A_i(x)\rightarrow -A_i({\cal T} x)\qquad ~~~i=1,2,3\\
 \phi^A(x)\rightarrow  \phi^A({\cal T} x)\qquad~~~ A=4,5,6\\
 \phi^{\dot A}(x)\rightarrow  -\phi^{\dot A}({\cal T} x)\qquad ~~\dot A=7,8,9\\
\lambda(x)\rightarrow\Gamma^{0456}\, \lambda({\cal T} x)\,,
 \end{cases}
 \label{Tact}
 \eeq
  where 
  \beq
  {\cal T} x\equiv (-x^0,x^1,x^2,x^3)\, 
 \eeq
 reverses  the time coordinate. $\phi^A$ are time-reversal scalars  while $\phi^{\dot A}$ are pseudo-scalars.

 Invariance of the dimensionally reduced ${\cal N}=4$ SYM Lagrangian \eqref{tendsym} under \eqref{Tact}  follows 
from  the antilinearly  of ${\mathsf T}$,  so that  ${\mathsf T} a {\mathsf T}^{-1}=a^*$ for $a\in\mathbb C$, and the fact that     $B\equiv\Gamma^{0456}$ obeys the  following   equations, required for the Lagrangian to be ${\mathsf T}$-invariant:
 \beq
 \begin{aligned}
 B^TB&=1\,,\\
 B^T\Gamma^{0i} B&=- \Gamma^{0i}\,,\\
  B^T\Gamma^{0A} B&=  \Gamma^{0A}\,,\\
  B^T\Gamma^{0{\dot A}} B&= - \Gamma^{0{\dot A}}\,,
 \end{aligned}
 \eeq
  where we have used    that $\Gamma^M$ are real.
 The  transformation (\ref{Tact})  defines a
    time-reversal    symmetry for an arbitrary Yang-Mills coupling $g^2$.
 
 The topological terms in the action are reversed under time-reversal symmetry. For instance, $\mathsf{T}$ flips the sign of the topological term \eqref{insta}, so that under time-reversal 
\begin{equation}
\mathsf{T}: \tau \mapsto -\bar{\tau}\,.
\label{Tontau}
\end{equation}
  The theory is $\mathsf{T}$-invariant -- analogously to the case of parity -- when the continuous and discrete theta angles are chosen such that their contributions to the path integral evaluate to $\pm 1$.

 The time-reversal transformation \eqref{Tact} defines a $\mathrm{Pin}^+$ structure since
 \beq
  \mathsf T^2= (-1)^F\,.
 \eeq
This implies that ${\cal N}=4$ SYM can be consistently formulated on an  unorientable  manifold   admitting a $\mathrm{Pin}^+$ structure~\cite{Kapustin:2014dxa,Witten:2015aba}. Any other choice of time-reversal   is obtained  from (\ref{Tact}) by combining it with a transformation belonging to the connected component of the Lorentz group acting on 
 $\mathbb R^{1,3}\times \mathbb R^{6}$. Doing this, it is possible\footnote{For example, by composing \eqref{Tact} with a   $\mathrm{Spin}(6)$ transformation, we can define a   time-reversal symmetry $\widetilde{\mathsf{T}}$ such that $\widetilde{\mathsf{T}}: \lambda(x) \rightarrow \Gamma^{04} \lambda({\cal T}x)$.} to define a time-reversal symmetry transformation $\widetilde{\mathsf{T}}$ with 
 \beq
 \widetilde{\mathsf{T}}^2 = 1\,.
 \eeq
Using this   time-reversal symmetry, ${\cal N}=4$ SYM can   be defined on an unoriented  manifold   admitting a $\mathrm{Pin}^-$ structure~\cite{Kapustin:2014dxa,Witten:2015aba}.

 We  turn now to the action of the time-reversal on local, line, and surface operators in \( \mathcal{N}=4 \) SYM, focusing on  \( \mathfrak{g} = \mathfrak{su}(N) \).

 \medskip
 \noindent
 $\bullet$ Local operators
  \medskip
  
  The analysis here closely parallels that of the parity transformation. As in that case, chiral primary operators transforming in the $(0, \Delta, 0)$ representation of $\mathrm{Spin}(6)$ decompose into operators $\mathcal{O}_\Delta^{j_1, j_2}$, each transforming in a spin $(j_1, j_2)$ representation of $\mathrm{Spin}(3) \times \mathrm{Spin}(3)$. Under the time-reversal symmetry \eqref{Tact}, these operators transform as
\begin{equation}
\mathsf{T} \left( \mathcal{O}_\Delta^{j_1, j_2}(x) \right) = (-1)^{j_2} \, \mathcal{O}_\Delta^{j_1, j_2}(\mathcal{T}x) \,.
\label{TonCPO}
\end{equation}
As a result, there exist $\mathsf{T}$-odd operators for every value of $\Delta$.

 \medskip
 \noindent
 $\bullet$ Line operators
 \medskip

There are two qualitatively distinct types of line operators to consider, distinguished by how they are embedded in \( \mathbb{R}^{1,3} \). Without loss of generality, the operator can be taken to   lie along the spatial direction\footnote{Any line operator lying in the fixed locus \( \mathbb{R}^3 \) at \( x^0 = 0 \) behaves identically under \( \mathsf{T} \).} \( x^3 \), or along the temporal direction \( x^0 \), which is inverted by time-reversal. We distinguish the latter case by placing a hat on the operator and consider those that transform simply under time-reversal.

\smallskip
 
\noindent
\textit{Wilson lines:}  The  time-reversal action on a Wilson line is most clearly obtained 
from the action of  \( \mathsf{T} \)  on the field produced by the line. For a Wilson line extending along $x^3$
 and coupling to a scalar $\phi\in \phi^{\dot A}$
 \begin{equation}
\mathsf{T}:
\begin{cases}
F_{30}(x) \rightarrow F_{30}(\mathcal{T}x) = -F_{30}(x) \,, \\
F_{31}(x) \rightarrow  -F_{31}(\mathcal{T}x) = -F_{31}(x) \,, \\
F_{32}(x) \rightarrow  - F_{32}(\mathcal{T}x) = -F_{32}(x) \,, \\
\phi(x) \rightarrow - \phi(\mathcal{T}x) = - \phi(x) \,.
\end{cases}
\end{equation}
In the last equal sign we have used the field produced by the line defect.\footnote{The chromoelectric field produced by a Wilson line along coordinate $x^a$ is $F_{ai}\propto {x^i\over |x|^3}$, where $x^i$ are the remaining three coordinates, and $\phi\propto {1\over |x|}$.} 
For a line operator along $x^0$ and with $\phi\in \phi^{A}$
\begin{equation}
\mathsf{T}:
\begin{cases}
F_{01}(x) \rightarrow F_{01}(\mathcal{T}x) = F_{01}(x) \,, \\
F_{02}(x) \rightarrow  F_{02}(\mathcal{T}x) = F_{02}(x) \,, \\
F_{03}(x) \rightarrow   F_{03}(\mathcal{T}x) = F_{03}(x) \,, \\
\phi(x) \rightarrow  \phi(\mathcal{T}x) =  \phi(x) \,.
\end{cases}
\end{equation}
 Therefore, the transformation of Wilson  line operators under \( \mathsf{T} \) is given by:
\begin{equation}
\begin{aligned}
\mathsf{T}(W_{R}( \phi^{\dot A})) &= W_{\overline{R}}(\phi^{\dot A}) \,, \\
\mathsf{T}(\hat W_{R}(\phi^{ A})) &= \hat W_{{R}}(\phi^{  A}) \,,
\end{aligned}
\end{equation}
while its transformation under \( \mathsf{C}\mathsf{T} \) is:
\begin{equation}
\begin{aligned}
\mathsf{C}\mathsf{T}(W_{R}(\phi^{\dot A})) &= W_{R}(\phi^{\dot A}) \,, \\
\mathsf{C}\mathsf{T}(\hat W_{R}(\phi^{A})) &= \hat W_{\overline{R}}(\phi^{A}) \,.
\end{aligned}
\end{equation}

\smallskip
 
\noindent
\textit{'t Hooft lines:} 

The time-reversal  properties of 't Hooft line operators are determined by analyzing how the singularity in equation~\eqref{thooftsingu} transforms under the action of \( \mathsf{T} \). For a line operator extending along the \( x^3 \) direction and coupled to a scalar \( \phi \in \phi^{A} \), the parity transformation acts as (cf. \eqref{thooftsingu})
 \beq
 \mathsf{T}:
\begin{cases}
F_{01}(x) \rightarrow F_{01}(\mathcal{T}x) = F_{01}(x) \,, \\
F_{02}(x) \rightarrow  F_{02}(\mathcal{T}x) = F_{02}(x) \,, \\
F_{12}(x) \rightarrow  - F_{12}(\mathcal{T}x) = F_{12}(x) \,, \\
\phi(x) \rightarrow  \phi(\mathcal{T}x) =  \phi(x) \,.
\end{cases}
\eeq
For a line operator along $x^0$ and with $\phi\in \phi^{\dot A}$
\beq
 {\mathsf T}:
  \begin{cases}
F_{ij}(x)\rightarrow - F_{ij}({\cal T}x)=-F_{ij}(x)\qquad i,j=1,2,3\\
\phi(x)\rightarrow -\phi({\cal T}x)= -\phi(x)\,.
\end{cases}
\eeq
 Therefore, the transformation of 't Hooft line operators under \( \mathsf{T} \) is given by:
\begin{equation}
\begin{aligned}
\mathsf{T}(T_{R}(\phi^{A})) &= T_{{R}}(\phi^{A}) \,, \\
\mathsf{T}(\hat{T}_{R}(\phi^{\dot{A}})) &= \hat{T}_{\overline{R}}(\phi^{\dot{A}}) \,,
\end{aligned}
\end{equation}
while under \( \mathsf{C}\mathsf{P} \), we have:
\begin{equation}
\begin{aligned}
\mathsf{C}\mathsf{T}(T_{R}(\phi^{A})) &= T_{\overline{R}}(\phi^{A}) \,, \\
\mathsf{C}\mathsf{T}(\hat{T}_{R}(\phi^{\dot{A}})) &= \hat{T}_{{R}}(\phi^{\dot{A}}) \,.
\end{aligned}
\end{equation}

This implies, using the  results summarized in Table~\ref{WilsonTCT}, that  $\mathsf{T}$ and $\mathsf{CT}$ are 
exchanged by $S$-duality 
\beq
  S\,\mathcal{R} : \quad \mathsf{T} \longleftrightarrow \mathsf{CT}\,,
  \label{TtoCT}
\eeq
or more  explicitly:
\beq
(S\,\mathcal{R})  \, \mathsf{T}\,  (S\,\mathcal{R} )^{-1}=\mathsf{CT}\,.
  \label{CtoCTex}
\eeq

\begin{table}[h!]
\centering
\renewcommand{\arraystretch}{1.4}
\begin{tabular}{|c|c|c|}
\hline
\text{Line Operator} & \textbf{  \( \mathsf{T} \)} & \textbf{  \( \mathsf{CT} \)} \\
\hline
\( W_{R}(\phi^{\dot{A}}) \) & \( W_{\overline{R}}(\phi^{\dot{A}}) \) & \( W_{R}(\phi^{\dot{A}}) \) \\
\hline
\( T_{R}(\phi^{A}) \) & \( T_{R}(\phi^{A}) \) & \( T_{\overline{R}}(\phi^{A}) \) \\
\hline
\( \hat{W}_{R}(\phi^{A}) \) & \( \hat{W}_{R}(\phi^{A}) \) & \( \hat{W}_{\overline{R}}(\phi^{A}) \) \\
\hline
\( \hat{T}_{R}(\phi^{\dot{A}}) \) & \( \hat{T}_{\overline{R}}(\phi^{\dot{A}}) \) & \( \hat{T}_{R}(\phi^{\dot{A}}) \) \\
\hline
\end{tabular}
\caption{Transformation of Wilson and 't Hooft lines under \( \mathsf{T} \) and \( \mathsf{CT} \).}
\label{WilsonTCT}
\end{table}

 \medskip
 \noindent
 $\bullet$ Surface  operators
 \medskip

Surface operators supported on the two-dimensional plane $\Sigma = \mathbb{R}^{1,1}$, with local coordinates $(x^0, x^3)$, admit a simple action under the time-reversal symmetry $\mathsf{T}$. Each operator couples to a complex scalar field as per \eqref{surface1}, where one chooses either
\beq
  \Phi \in \phi^A
  \quad\text{or}\quad
  \Phi \in \phi^{\dot A}.
\eeq

 Each operator couples to a complex scalar field, where one chooses either
\beq
  \Phi \in \phi^A
  \quad\text{or}\quad
  \Phi \in \phi^{\dot A}.
\eeq
We attach to each surface operator a label indicating which complex scalar it couples to.

The time-reversal  properties of   surface operators follow from how the singularity in \eqref{surface1} and the defect insertion in \eqref{surface2} transform under the action of \(\mathsf{T}\) in \eqref{Tact}.  One finds that under \(\mathsf{T}\):\footnote{Under $\mathsf T$, the electric field $F_{03}$ along $\mathbb R^{1,1}$ is even while the magnetic field  $F_{12}$ is odd.}
\begin{equation}
\begin{aligned}
\mathsf{T}\bigl(\mathcal{O}_\Sigma(\alpha,\beta,\gamma,\eta)[\phi^{A}]\bigr)
  &= \mathcal{O}_\Sigma(-\alpha,\beta,\gamma,\eta)[\phi^{ A}], \\[6pt]
\mathsf{T}\bigl(\mathcal{O}_\Sigma(\alpha,\beta,\gamma,\eta)[\phi^{\dot A}]\bigr)
  &= \mathcal{O}_\Sigma(-\alpha,-\beta,-\gamma,\eta)[\phi^{\dot A}],
\end{aligned}
\label{actonsurfttt}
\end{equation}
and under \(\mathsf{CT}\):
\begin{equation}
\begin{aligned}
\mathsf{CT}\bigl(\mathcal{O}_\Sigma(\alpha,\beta,\gamma,\eta)[\phi^{ A}]\bigr)
  &= \mathcal{O}_\Sigma(\alpha,-\beta,-\gamma,-\eta)[\phi^{ A}], \\[6pt]
\mathsf{CT}\bigl(\mathcal{O}_\Sigma(\alpha,\beta,\gamma,\eta)[\phi^{\dot A}]\bigr)
  &= \mathcal{O}_\Sigma(\alpha,\beta,\gamma,-\eta)[\phi^{\dot A}].
\end{aligned}
\end{equation}

\begin{table}[h!]
\centering
\renewcommand{\arraystretch}{1.4}
\begin{tabular}{|c|c|c|}
\hline
\text{Surface Operator} & \textbf{\( \mathsf{T} \)} & \textbf{\( \mathsf{CT} \)} \\
\hline
\( \mathcal{O}_{\Sigma}(\alpha,\beta,\gamma,\eta)[\phi^{A}] \) 
& \( \mathcal{O}_{\Sigma}(-\alpha,\beta,\gamma,\eta)[\phi^{A}] \) 
& \( \mathcal{O}_{\Sigma}(\alpha,-\beta,-\gamma,-\eta)[\phi^{A}] \) \\
\hline
\( \mathcal{O}_{\Sigma}(\alpha,\beta,\gamma,\eta)[\phi^{\dot{A}}] \) 
& \( \mathcal{O}_{\Sigma}(-\alpha,-\beta,-\gamma,\eta)[\phi^{\dot{A}}] \) 
& \( \mathcal{O}_{\Sigma}(\alpha,\beta,\gamma,-\eta)[\phi^{\dot{A}}] \) \\
\hline
\end{tabular}
\caption{Transformation of surface operators  under \( \mathsf{T} \) and \( \mathsf{CT} \).}
\label{surfaceTCT}
\end{table}

Using the action of $S$   in $\Gamma\subset SL(2,\mathbb{Z})$ on the surface operator parameters \eqref{actonsurf},
and  the  results summarized in Table~\ref{surfaceTCT}, we find that indeed $\mathsf{T}$ and $\mathsf{CT}$ are 
exchanged by $S$-duality as in~\eqref{CtoCTex} and  that:
\beq
(S\,\mathcal{R})  \mathsf{T} (S\,\mathcal{R} )^{-1}=\mathsf{CT}\,.
\eeq

\subsection{$\mathsf{CPT}$-symmetry}
\label{sec:CPTt}

A canonical, antilinear, antiunitary $\mathsf{CPT}$ symmetry transformation can be defined in any relativistic quantum field theory. Even if the theory possesses additional 0-form symmetries, that could be composed with a fiducial $\mathsf{CPT}$ transformation, a canonical choice is singled out by deforming the theory in a Lorentz-invariant fashion with Hermitean operators\footnote{$\mathsf{CPT}$ symmetry is only guaranteed in a Lorentz invariant theory with a Hermitean action. For example, $\hbox{Tr}\left(F_{\mu\nu} F^{\nu\rho}F_{\rho}^{\ \mu}  \right)$ is Lorentz invariant but not  $\mathsf{CPT}$-invariant, as the operator is not real.}
 such that all symmetries are broken.\footnote{Higher form symmetries,  while unbroken by deformations,   do not the modify the action of  $\mathsf{CPT}$.} In  the   deformed theory with no symmetries, the $\mathsf{CPT}$ transformation becomes essentially universal, as it can only differ from a fiducial  transformation by the action of $(-1)^F\in\mathrm{Spin}(1,3)$~\cite{Hason:2020yqf,Witten:2025ayw}, which is an unbreakable symmetry in a Lorentz-invariant theory. Once the choice of which odd number of spatial coordinates are reflected is fixed, just one coordinate in this paper, the $\mathsf{CPT}$ transformation is fully determined up to $(-1)^F$. With our choice of coordinate reflections, the canonical $\mathsf{CPT}$ transformation can be understood as the Wick rotation of a   $\pi$ rotation  in the $x^0_{\rm E} - x^3$ plane in $\mathbb{R}^4$.

In $\mathfrak{su}(N)$ ${\cal N}=4$ SYM, $\mathsf{CPT}$ acts by
  \beq
 \mathsf{CPT}:
 \begin{cases}
 A_0(x)\rightarrow  -A_0^T({\cal P}{\cal T} x)\qquad \\
  A_i(x)\rightarrow   A^T_i({\cal P}{\cal T} x) \qquad ~~~i=1,2\\
  A_3(x)\rightarrow   -A_3^T({\cal P}{\cal T} x) \qquad \\
 \phi^I(x)\rightarrow  (\phi^I)^T({\cal P}{\cal T} x) \\
\lambda(x)\rightarrow   \Gamma^{03}\, \lambda^T({\cal P}{\cal T} x)\,,
 \end{cases}
 \label{CPTact}
 \eeq
  where 
  \beq
  {\cal P}{\cal T}  x\equiv (-x^0,x^1,x^2,-x^3)\, 
 \eeq
 reverses  the time coordinate $x^0$ and the spatial  coordinate $x^3$. It obeys $\mathsf{CPT}^2=1$.
It is a symmetry of ${\cal N}=4$ for arbitrary complexified coupling, so that
\begin{equation}
\mathsf{CPT}:   \tau \rightarrow  \tau \,,
\label{CPTontau}
\end{equation}
   and    for arbitrary discrete theta angles.

A Wilson and 't Hooft line coupled to an arbitrary scalar   in the plane fixed by ${\cal P}{\cal T}$, say along $x^1$,  
 transform as:
\begin{equation}
\begin{aligned}
\mathsf{CPT}(W_{R})  &= W_{R} \,, \\
\mathsf{CPT}(T_{R})  &= T_{R} \,.
\label{CPTW1}
\end{aligned}
\end{equation}
A Wilson and 't Hooft loop along   $x^0$ or $x^3$, and not sourcing a scalar, transform   under $\mathsf{CPT}$ as:
\begin{equation}
\begin{aligned}
\mathsf{CPT}(\hat W_{R})  &= \hat W_{\overline{R}} \,, \\
\mathsf{CPT}(\hat T_{R})  &= \hat T_{\overline{R}} \,.
\end{aligned}
\label{CPTW2}
\end{equation}

A surface operator  supported on   $\Sigma = \mathbb{R}^{1,1}$, with local coordinates $(x^0, x^3)$, coupled to an
arbitrary complex scalar field obeys:\footnote{Under $\mathsf{CPT}$, the electric field $F_{03}$ along $\mathbb R^{1,1}$  and magnetic field    $F_{12}$ are even.}
\begin{equation}
\begin{aligned}
\mathsf{CPT}\bigl(\mathcal{O}_\Sigma(\alpha,\beta,\gamma,\eta)\bigr)
   = \mathcal{O}_\Sigma(\alpha,\beta,\gamma,\eta)\,.
 \end{aligned}
 \label{CPTS1}
\end{equation}

All this shows that $\mathsf{CPT}$ is invariant under $S$-duality:
\beq
(S\,\mathcal{R}) \, \mathsf{CPT} \, (S\,\mathcal{R} )^{-1}=\mathsf{CPT}\,.
  \label{CtoC}
\eeq

\section{Holographic Dual of $\mathsf{C}$-$\mathsf{P}$-${\mathsf T}$ Symmetry in ${\cal N}=4$ SYM}
\label{sec:CPTads}

Type IIB string theory, being chiral and having spacetime fermions, requires the ten-dimensional spacetime manifold   to be both oriented and admit a spin structure.
The fact that Type IIB string theory cannot be consistently defined on an unorientable manifold,\footnote{This stands in contrast to M-theory, which can be formulated on an unorientable manifold that admits a \(\textrm{Pin}^+\) structure \cite{Witten:1996md,Witten:2016cio}.}
    informs our search for   the dual   of the \(\mathsf{C}\), \(\mathsf{P}\), and \(\mathsf{T}\) symmetries of \({\cal N}=4\) SYM discussed in section \ref{sec:CPT}. The candidate transformations must be orientation preserving symmetries of  Type IIB string theory in the  \(AdS_5 \times S^5\) background.

We   turn to the worldsheet symmetries of Type IIB string theory that can be preserved at arbitrary values of the string coupling $g_s$. These symmetries are   ingredients for the holographic duals of the \(\mathsf{C}\), \(\mathsf{P}\), and \(\mathsf{T}\) symmetries of \({\cal N}=4\) SYM, which persist at all values of the gauge coupling since $g^2=4\pi g_s$.
 These include the worldsheet symmetries:    
\beq
\begin{aligned}
\Omega &: \text{worldsheet parity}\,, \\[+2pt]
(-1)^{F_L} &: \text{spacetime fermion number parity from  left-moving sector of string worldsheet}\,.
\end{aligned}
\eeq
The   generators $a = \Omega\, (-1)^{F_L}, \,  b = \Omega$ obey the group relations~\cite{Dabholkar:1997zd}
\beq
a^4 = b^2 = 1, \quad bab = a^{-1}.
\eeq
Therefore, the potential  symmetry group of Type IIB string theory for any string coupling 
is the dihedral symmetry group of eight elements.\footnote{\( \Omega\, (-1)^{F_L} \) is of order 4 and obeys
$
\left(\Omega\, (-1)^{F_L}\right)^2 = (-1)^F \equiv (-1)^{F_L} (-1)^{F_R}\,,
$ where $(-1)^F$ denotes $10d$ spacetime fermion parity.
}
  Of course, in a given Type IIB background, such as the $AdS_5\times S^5$, only a subset of these symmetries,
    combined with spacetime transformations,  is preserved.

The action of $\Omega$, $(-1)^{F_L}$ and $\Omega(-1)^{F_L}$ on the massless bosonic fields of Type IIB string theory is (see e.g. \cite{Dabholkar:1997zd}):
\beq
\Omega:
\begin{cases}
\tau\rightarrow -\bar \tau\,,\\
B_2\rightarrow -B_2\,,\\
C_2\rightarrow C_2\,,\\
C_4\rightarrow-C_4\,,
\end{cases}
~~~
(-1)^{F_L}:
\begin{cases}
\tau\rightarrow -\bar \tau\,,\\
B_2\rightarrow B_2\,,\\
C_2\rightarrow -C_2\,,\\
C_4\rightarrow-C_4\,,
\end{cases}
~~~
\Omega(-1)^{F_L}:
\begin{cases}
\tau\rightarrow   \tau\,,\\
B_2\rightarrow -B_2\,,\\
C_2\rightarrow -C_2\,,\\
C_4\rightarrow  C_4\,.
\end{cases}
\label{transffields}
\eeq
$C_p$ are the RR $p$-forms, $B_2$ the NSNS two-form and 
\beq
 \tau=C_0+ i e^{-\phi}\,,
 \label{tauiib} 
 \eeq
with $\phi$ the dilaton. 
The metric is   invariant. The action of these symmetries on D-branes, F1-string and NS5-branes follows from~\eqref{transffields}.  We employ the same symbol $\tau$ for the Type  IIB axio-dilaton in~\eqref{tauiib} and for the complexified coupling of $\mathcal{N}=4$ SYM in~\eqref{tauSYM}, because the AdS/CFT dictionary identifies them through
\beq
{\theta\over 2\pi}+{4\pi   i\over g^2}=C_0+ i e^{-\phi}\,.
\eeq
 In this section $\tau$ always refers to \eqref{tauiib}.
 
The \(S\)-transformation in the nonperturbative duality group of Type IIB string theory,\footnote{The full nonperturbative duality group of Type IIB string theory, when the action on fermions is properly accounted for, is a \(\mathrm{Pin}^+\) double cover of \(GL(2,\mathbb{Z})\)~\cite{Tachikawa:2018njr}. The perturbative symmetries \(\Omega\)  and \((-1)^{F_L}\)   extend the familiar \(SL(2,\mathbb{Z})\) S-duality group -- valid when fermions are ignored -- to the larger group \(GL(2,\mathbb{Z})\). In the full theory, this extension lifts to a \(\mathrm{Pin}^+\) structure over \(GL(2,\mathbb{Z})\), capturing the correct transformation properties of fermionic states (see also \cite{Debray:2023yrs}).}
exchanges  \(\Omega\) and \((-1)^{F_L}\)~\cite{Sen:1996na}:
\beq
\Omega \; \overset{\text{S}}\longleftrightarrow\; (-1)^{F_L}\,,
\eeq
while their product \(\Omega (-1)^{F_L}\) remains invariant under the action of $S$.

 \smallskip
 We  now  recall the Type IIB  $AdS_5\times S^5$ bulk description of the gauge  theory operators we have discussed  in section \ref{sec:CPT}:
 
 \medskip
 \noindent
 $\bullet$ Local operators: The chiral primary operator \({\cal O}_\Delta\) corresponds to a scalar field \(s_\Delta\) in Type IIB supergravity on \(AdS_5 \times S^5\)~\cite{Witten:1998qj}. This field is  a linear combination of fluctuations of the $10d$  metric and the RR four-form \(C_4\) along the \(S^5\), with a spherical harmonic as its wavefunction~\cite{Kim:1985ez}. The quantum numbers of this spherical harmonic determine both the \(\mathrm{Spin}(6)\)   representation and the scaling dimension \(\Delta\) of the dual operator \({\cal O}_\Delta\):
 \beq
\begin{aligned}
 {\cal O}_\Delta  \leftrightarrow  s_\Delta 
 \label{CPOsugra}
\end{aligned}
\eeq
 The rest of the supergravity modes sit in a short  multiplet with $s_\Delta$ being the bottom component.
 
  \medskip
 \noindent
 $\bullet$ Line operators: The bulk description of a Wilson \eqref{Wilsonf} and 't Hooft line \eqref{thooftsingu} depends on the choice representation $R$.  To describe it, it is convenient to foliate  $AdS_5$ by $AdS_2\times S^2$ slices and $S^5$ by $S^4$ slices, the latter being:
 \beq
 d\Omega_5^2=d\theta^2+\sin^2\theta d\Omega_4^2 ~~~\qquad 0\leq \theta\leq \pi\,.
 \label{spherefour}
 \eeq
   We refer to $\theta$ as the latitude angle of the $S^4$.
   
   A Wilson line  in the fundamental representation is described by a F1-string~\cite{Maldacena:1998im,Rey:1998ik},
   a D5$_k$-brane describes  the $k$-th antisymmetric~\cite{Yamaguchi:2006tq,Gomis:2006sb,Gomis:2006im}  and a D3$_k$-brane describes  the $k$-th symmetric representation~\cite{Gomis:2006sb,Gomis:2006im}(see also \cite{Drukker:2005kx}). The index $k$ refers to the amount of F1-string charge dissolved on the brane,
   induced by turning on an electric field along the worldvolume. For a 't   Hooft line,  it is    the $S$-dual brane, and $k$ denotes the
   D1-brane charge.
  In what follows,   $\overline{\text{brane}}\equiv \text{antibrane}$, that is a brane with its orientation reversed. A minus sign in the subindex  $\text{brane}_{-k}$ signals  that   the brane carries opposite F1-string (or D1-string) charge  compared to $\text{brane}_{k}$.
  The bulk realization of Wilson lines is:
 \beq
\begin{aligned}
 W_{\text{fund}}  &\leftrightarrow \text{F1 on}~ AdS_2~\text{at}~\theta=0\,,\\
 W_{ \overline {\text{fund}}}&\leftrightarrow \overline{{\text F1}}~ \text{on}~ AdS_2~\text{at}~\theta=\pi\,, \\
W_{\text{A}_k}&\leftrightarrow \text{D5}_k~ \text{on}~ AdS_2\times S^4~\text{with}~S^4\subset S^5~ \text{latitude angle}~  \theta_k\,,  \\  
 W_{\text{S}_k}  & \leftrightarrow \text{D3}_k~ \text{on}~ AdS_2\times S^2 \subset AdS_5~\text{at}~\theta=0\,,\\
 W_{\overline {\text{S}_k}}  & \leftrightarrow \text{D3}_{-k}~ \text{on}~ AdS_2\times S^2 \subset AdS_5~\text{at}~\theta=\pi\,.
\end{aligned}
\label{Wilsonmap}
\eeq
The brane describing a Wilson line in the $\overline {\text{fund}}$ and $\overline {\text{S}_k}$ sits at the antipodal point on $S^5$ compared to  the brane describing a Wilson line in representations $\text{fund}$ and $\text{S}_k$. Indeed, this reverses the sign of the coupling of the Wilson  and 't Hooft line to the scalar field (cf.  \eqref{Wilsonf}\eqref{thooftsingu}). The latitude angle $\theta_k$ of the $S^4\subset S^5$ is determined by the amount of F1 (or D1) charge $k$~\cite{Pawelczyk:2000hy,Camino:2001at,Yamaguchi:2006tq}. It is   bounded by the compactness of $S^5$ to $0\leq k\leq N$.

For 't Hooft lines: 
 \beq
\begin{aligned}
 T_{\text{fund}}  &\leftrightarrow \text{D1 on}~ AdS_2~\text{at}~\theta=0\,, \\
T_{ \overline {\text{fund}}}&\leftrightarrow \overline{{\text D1}}~ \text{on}~ AdS_2~\text{at}~\theta=\pi\,, \\
T_{\text{A}_k}&\leftrightarrow \text{NS5}_k~ \text{on}~ AdS_2\times S^4~\text{with}~S^4\subset S^5~ \text{latitude angle}~  \theta_k\,,  \\  
T_{\text{S}_k}  & \leftrightarrow \text{D3}_k~ \text{on}~ AdS_2\times S^2 \subset AdS_5~\text{at}~\theta=0\,,\\
T_{\overline{\text{S}}_k}  & \leftrightarrow \text{D3}_{-k}~ \text{on}~ AdS_2\times S^2 \subset AdS_5~\text{at}~\theta=\pi\,.
\label{hooftmap}
\end{aligned}
\eeq
 The $\text{NS5}_k$ and $\text{D3}_k$ branes carry now  ${\text D1}$-string charge. This is induced by turning
on    a worldvolume electric field on $AdS_2$ on $\text{NS5}_k$ and a worldvolume magnetic field on $S^2$   on $\text{D3}_k$ respectively.  We will frequently refer to \eqref{Wilsonmap} and \eqref{hooftmap} in analyzing how symmetries act on branes 
and in identifying their counterparts in \({\cal N}=4\) SYM.

A Wilson line    in a large representation $R$ admits a bulk description in terms of an asymptotically $AdS_5\times S^5$, smooth, topologically rich bubbling solution of Type IIB string theory~ \cite{Yamaguchi:2006te,Lunin:2006xr,DHoker:2007mci}. The solution is  a  nontrivial fibration of $AdS_2\times S^2\times S^4$ over the upper half plane. The choice of representation $R$ of Wilson line is encoded in   a choice of  ``coloring" of the real axis, which specifies the locus at which $S^2$ and $S^4$ shrink in a smooth fashion.\footnote{$S^2$ shrinks on black and $S^4$ on white.}  The coloring describing a Wilson line in a representation $R$ is 
obtained by describing $R$ by its  Russian  Young diagram and projecting it onto the real axis, drawing a black interval for every boundary of the tableau in the south-east direction, and white otherwise~\cite{Yamaguchi:2006te,Gomis:2008qa}(see also \cite{Gomis:2006mv,Gomis:2007kz}), as in    figure \ref{fig:bubblingw}:
  \begin{figure}[H]
    \centering
    \includegraphics[width=0.65\textwidth]{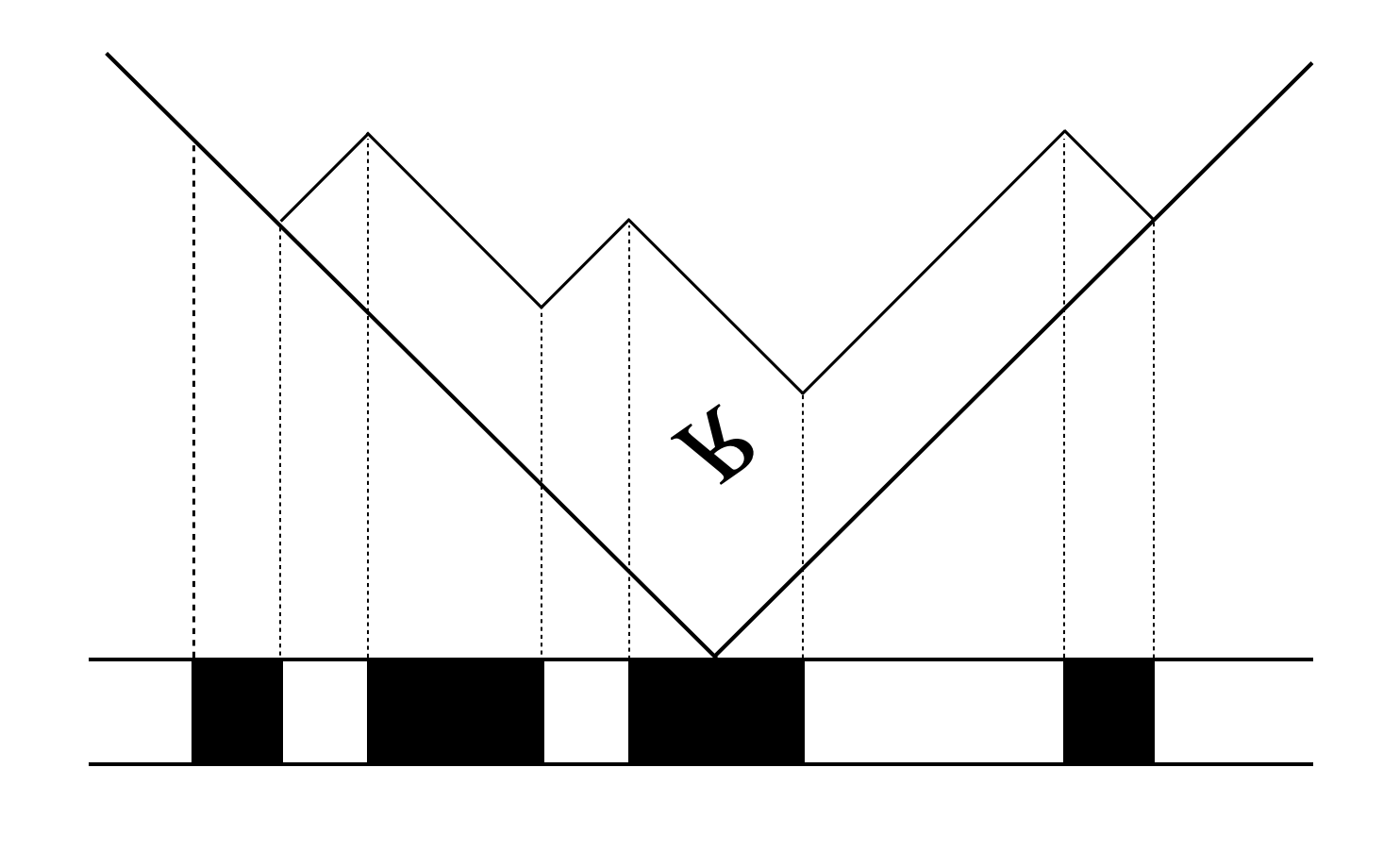}
    \caption{Map between choice of representation $R$ and boundary conditions on the real axis. The thicker leftmost vertical line bounds the number columns in tableau by $N$.}
    \label{fig:bubblingw}
  \end{figure}

  \medskip
 \noindent
 $\bullet$ Surface operators: A surface operator ${\cal O}_{\Sigma}(\alpha,\beta,\gamma,\eta)$ with a Levi group $\mathbb L$ and Lie algebra $\mathfrak{su}(N-1)\times\mathfrak{u}(1)$ is described by a single D3-brane in $AdS_5\times S^5$~\cite{Gukov:2006jk}.   Since  ${\cal O}_{\Sigma}(\alpha,\beta,\gamma,\eta)$ couples to a complex scalar,    the dual D3-brane embedding is simplest in an  $AdS_3\times S^1$ foliation of $AdS_5$
  \beq
 ds^2_{AdS_5}=d\rho^2+\cosh^2\rho \,ds^2_{AdS_3}+ \sinh^2\rho\, d\psi^2  \,.
 \label{AdS}
 \eeq
 and an $S^3\times S^1$ foliation of $S^5$
 \beq
 d\Omega_5^2=d\vartheta^2+\cos^2\vartheta \,  d \phi^2+ \sin^2\vartheta \, d\Omega_3^2 ~~~\qquad 0\leq \vartheta\leq \pi/2\,.
 \label{S5met}
 \eeq
The bulk description of this surface operator is in terms of following D3-brane~\cite{Gukov:2006jk} (see also \cite{Constable:2002xt,Drukker:2008wr}):
 \beq
\begin{aligned}
 {\cal O}_{\Sigma}(\alpha,\beta,\gamma,\eta)  &\leftrightarrow \text{D3 on}~ AdS_3\times S^1~~~\text{with} ~\rho=\rho_0\,,  \psi+\phi=\phi_0 ~\text{at}~\vartheta=0\,.
\end{aligned}
\label{surfaceD3}
\eeq
The parameter mapping is~\cite{Gukov:2006jk,Drukker:2008wr} 
\beq
\alpha=\oint_{S^1} {a\over 2\pi}\,,\qquad \eta=\oint_{S^1} {\tilde a\over 2\pi}\,,\qquad \beta+i\gamma={\sqrt{\lambda}\over 2\pi}\sinh \rho_0\, e^{i\phi_0}\,,
\label{surfacemap}
\eeq
where $a$ and $\tilde a$ are the $U(1)$ gauge field and dual gauge field on the D3-brane worldvolume. 
\begin{figure}[H]
    \centering
    \includegraphics[width=0.35\textwidth]{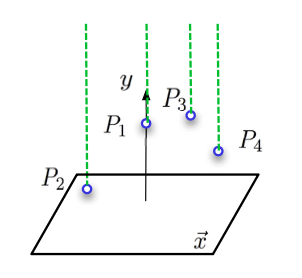}
    \caption{Data determining the  solution dual to a general surface operator ${\cal O}_{\Sigma}(\alpha,\beta,\gamma,\eta)$.
    The $AdS_5\times S^5$ corresponds to single point charge. The mapping of parameters is in   \eqref{mapsurfb}.}  
    \label{fig:bubblings}
  \end{figure}

A surface operator with a general  Levi group $\mathbb L$ with $M$ factors and Lie algebra $\prod_{l=1}^M S( \mathfrak{u}(N_l) )$
admits a bulk description in terms of an asymptotically $AdS_5\times S^5$, smooth, topologically rich bubbling solution of Type IIB string theory \cite{Gomis:2007fi} (see also \cite{Lin:2005nh}). The solution is  a  nontrivial fibration of $AdS_3\times S^1\times S^3$ over  $\mathbb R^3_+$. The solution is completely determined by $M+1$ point charges in $\mathbb R^3_+$   located at  $(\vec{x}_l, y_l)$ and nontrivial holonomies of $B_2$ and $C_2$ along nontrivial two-cycles $D_l$ (see figure
\ref{fig:bubblings}).\footnote{The $S^3$ shrinks smoothly at $y=0$ and the $S^1$ shrinks smoothly at the point sources. The  nontrivial two-cycle  $D_l$ is a line from infinity in $\mathbb R^3_+$ to the $l$-th point source.} The mapping of parameters is~\cite{Gomis:2007fi}:
\beq
N_l={y_l^2\over 4\pi l_p^4}\,,\qquad (\beta_l,\gamma_l)={\vec x_l\over 2\pi l_s^2}\,,\qquad \alpha_l=-\int_{D_l} {B_2\over 2\pi}\,,\qquad\eta_l=\int_{D_l} {C_2\over 2\pi}\,.
\label{mapsurfb}
\eeq

 \smallskip

In the remainder of this section, we identify the bulk counterparts of the $\mathsf{C}$, $\mathsf{P}$, and $\mathsf{T}$ symmetries of ${\cal N}=4$ SYM introduced in Section~\ref{sec:CPT}. Our strategy is to combine:
\begin{enumerate}
  \item   discrete symmetries of Type~IIB string theory reviewed above, with
  \item   discrete isometries of the $AdS_5 \times S^5$ background
\end{enumerate}
which are $10d$ spacetime orientation preserving symmetries of Type IIB string theory on $AdS_5 \times S^5$.
We  analyze the action of   bulk symmetries on the holographic description  of local operators, Wilson lines, 't~Hooft lines, and surface operators and show that they realize, in the bulk, the action of $\mathsf{C}$, $\mathsf{P}$, and $\mathsf{T}$ on all these gauge theory operators.

Each section begins with our proposed   bulk symmetry     dual to the corresponding discrete symmetry of ${\cal N}=4$ SYM.

\subsection{Holographic ${\mathsf C}$-symmetry}

 Charge conjugation symmetry \eqref{transfo} in \({\cal N}=4\) SYM is realized in Type IIB string theory in $AdS_5\times S^5$ by: 
 \beq
 \mathsf C\longleftrightarrow \Omega(-1)^{F_L}  R_{456789}\,.
 \eeq
 $R_I$ acts geometrically on $AdS_5\times S^5$ by reversing the $I$-th embedding coordinate
  \beq
 R_I: x^I\rightarrow -x^I 
 \eeq
defining the $S^5$
\beq
\sum_{I=4}^9 x^I x^I=1\,.
\eeq
Quick consistency checks  include that  the reflection \(R_{456789}\) preserves the orientation\footnote{Reflection of an even number of embedding coordinates on a sphere preserves its orientation, 
while reflection of an odd number reverses it. We will use this often henceforth.} of spacetime and\
that $\Omega(-1)^{F_L}  R_{456789}$ is a symmetry of the $AdS_5 \times S^5$ background, preserving both   the five-form flux and $\tau$, just as ${\mathsf C}$ does \eqref{Contau}.

As shown in section \ref{sec:C}, the  \(\mathsf{C}\)-symmetry  of \({\cal N}=4\) SYM commutes with the full 
\(PSU(2,2|4)\) superconformal symmetry. Consequently, its holographic dual -- namely the bulk transformation 
\(\Omega(-1)^{F_L} R_{456789}\) -- must preserve all the Killing spinors of the \( AdS_5 \times S^5 \) background. In the \( AdS_5 \times S^5 \)  coordinate system
 \beq
  ds^2_{AdS_5\times S^5}= X^2 \sum_{\mu=0}^3 \eta_{\mu \nu} dx^\mu dx^\nu+{1\over X^2}\sum_{I=4}^9 dx^I dx^I\,,
\label{metricads}
  \eeq
  where $X^2=\sum_{M=4}^9 x^I x^I$, 
 the complex, chiral \( AdS_5 \times S^5 \) Killing spinor,  in the obvious choice of frame,  is~\cite{Claus:1998yw}:
\beq
\epsilon(x)={\sqrt{X}}  (1+i \Gamma^{0123})\epsilon_s+ {\sqrt{X}}\left(- x^\mu \Gamma_\mu+{x^I \Gamma_I\over X^2} \right)
(1- i \Gamma^{0123})\epsilon_c \,.
\label{killspin}
\eeq
 $X$   the $AdS_5$ radial coordinate with boundary at $X\rightarrow \infty$. $\epsilon_s$ and $\epsilon_c$ are constant $10d$ Majorana-Weyl spinors that parametrize the sixteen Poincar\'e and conformal supersymmetries of ${\cal N}=4$ SYM at the $X\rightarrow \infty$ boundary.


  Under the    sphere antipodal map,  the \( AdS_5 \times S^5 \) Killing spinor \eqref{killspin} transforms as
\beq
R_{456789}: \epsilon(x)\rightarrow \Gamma^{456789} \epsilon((x^\mu,-x^I))= i \epsilon(x)\,, 
\eeq
where we have used that $\Gamma^{456789}$ commutes with $\Gamma^\mu$ and anticommutes with $\Gamma^I$, and that $\epsilon_s$ and $\epsilon_c$ have opposite $10d$ chirality: $\Gamma^{01\ldots 9}\epsilon_s=\epsilon_s$ and $\Gamma^{01\ldots 9}\epsilon_c=-\epsilon_c$. 
We also need to understand how $\Omega$ and $(-1)^{F_L}$ act on the Killing spinor $\epsilon(x)$.
The complex Killing spinor \eqref{killspin} is assembled from a pair of  Majorana-Weyl spinors of the same chirality coming from the left and right-movers on the string worldsheet:
\beq
\epsilon(x)=\epsilon_{\mathrm{r}}(x)+i\epsilon_{\mathrm{l}}(x)\,.
\eeq
 Since $\Omega$ acts as worldsheet parity and $(-1)^{F_L}$ reverses the sign of left-moving spacetime fermions, we have that:
 \beq
 \Omega: \epsilon_{\mathrm{r}}\leftrightarrow  \epsilon_{\mathrm{l}}\,, \qquad~~~
 (-1)^{F_L}:
 \begin{cases}
 \epsilon_{\mathrm{r}}\rightarrow \epsilon_{\mathrm{r}}\\
 \epsilon_{\mathrm{l}}\rightarrow -\epsilon_{\mathrm{l}}\,,
 \end{cases}
 \eeq
 and therefore 
 \beq
 \Omega(-1)^{F_L}: \epsilon(x)\rightarrow -i \epsilon(x)\,.
 \eeq
This analysis confirms that the combined transformation \(\Omega(-1)^{F_L} R_{456789}\) 
preserves all supersymmetries of \( AdS_5 \times S^5 \)
\beq
 \Omega(-1)^{F_L}R_{456789}\cdot \epsilon(x)=\epsilon(x)\,,
\eeq
 and therefore commutes with the full 
\(PSU(2,2|4)\) superconformal algebra -- just as charge conjugation does in \({\cal N}=4\) SYM.

 We now show that the bulk transformation \(\Omega(-1)^{F_L} R_{456789}\) acts on the supergravity duals of 
local, line, and surface operators exactly as the charge conjugation symmetry \(\mathsf{C}\) acts on the 
corresponding operators,   studied   in Section~\ref{sec:C}. Because \(\Omega(-1)^{F_L}\) leaves both the \(AdS_5 \times S^5\) metric and the RR four-form \(C_4\) 
unchanged, the only non-trivial effect of  \(\Omega(-1)^{F_L} R_{456789}\)  on the supergravity mode $s_\Delta$ comes from the spatial reflection \(R_{456789}\).  
This reflection flips the sign of any \(S^5\) spherical harmonic in the \((0,\Delta,0)\) representation of 
\(Spin(6)\) when \(\Delta\) is odd.  Consequently, under \(\Omega(-1)^{F_L} R_{456789}\)
\beq
 s_\Delta\rightarrow (-1)^\Delta s_\Delta\,,
\eeq
which realizes the transformation law \eqref{actonCPO} for the chiral primary operators 
\(\mathcal{O}_\Delta\) under \(\mathsf{C}\).

 We now introduce some useful notation. If a brane wraps a submanifold whose orientation is reversed under a symmetry transformation, we denote the resulting geometry with an overline. For example, \( \overline{S^4} \) denotes an $S^4$  with reversed orientation. Importantly, an antibrane wrapping an orientation-reversed geometry is physically equivalent to a brane wrapping the original, unreflected geometry. That is,
\begin{equation}
\overline{\text{brane}}~\text{on}~\overline{M} = \text{brane}~\text{on}~M\,.
\label{braneeq}
\end{equation}
We   use this notation  and equality throughout the remainder of this paper.

Let us turn to discussing Wilson lines.  Given that \(\Omega(-1)^{F_L}\) flips the sign of $B_2$ and $C_2$ and $R_{456789}$ is the antipodal map on $S^5$:
 \beq
 \begin{aligned}
  \text{F1}~ AdS_2~\text{at}~\theta=0 &\rightarrow \overline{{\text F1}}~  AdS_2~\text{at}~\theta=\pi\\ 
  \text{D1}~ AdS_2~\text{at}~\theta=0 &\rightarrow\overline{{\text D1}}~  AdS_2~\text{at}~\theta=\pi\,.
\end{aligned}
 \eeq
 This realizes ${\mathsf C}(W_{\text{fund}}) = W_{ \overline {\text{fund}}}$ and ${\mathsf C}(T_{\text{fund}}) = 
 T_{ \overline {\text{fund}}}$ in \eqref{Cwilson}\eqref{conT} using   \eqref{Wilsonmap}.
 
The analysis for higher representations involves several subtle points, which we now elucidate. Under the action of \(\Omega(-1)^{F_L} R_{456789}\):
  \beq
 \begin{aligned}
  \text{D5}_k~\text{on}~   AdS_2\times S^4~ \text{at}~\theta_k &\rightarrow \overline{\text{D5}}_{-k}~\text{on} ~ AdS_2\times \overline{S^4} ~ \text{at}~\pi-\theta_k=
  \text{D5}_{N-k} ~\text{on}~   AdS_2\times S^4 ~ \text{at}~\pi-\theta_k \\  
  \text{D3}_k  ~\text{on}~  AdS_2\times S^2 ~\text{at}~\theta=0  &\rightarrow
  \text{D3}_{-k}  ~\text{on}~  AdS_2\times S^2~\text{at}~\theta=\pi\,.
  \label{chargeCC}
\end{aligned}
 \eeq
 $\Omega(-1)^{F_L}$    maps a    D5-brane to a  $\overline{{\text D5}}$ and  $R_{456789}$
 maps  the original $S^4$ to the antipodal $S^4$,  and changes its orientation since all five embedding coordinates of $S^4$ are reflected. It also reverses its F1-string charge.  The F1-string charge dissolved on the D5-brane determines the latitude angle \( \theta_k \) of the embedded \( S^4 \subset S^5 \). In a gauge where the RR potential \( C_4 \) is regular near \( \theta = 0 \), the relation between the string charge \( k \) and the angle \( \theta_k \) is given by~\cite{Pawelczyk:2000hy,Camino:2001at,Yamaguchi:2006tq}
\begin{equation}
k = \frac{N}{\pi} \left( \theta_k - \frac{1}{2} \sin 2\theta_k \right),
\end{equation}
with \( 0 \leq k \leq N \). This relation implies the identity
\begin{equation}
\theta_{N - k} = \pi - \theta_k\,.
\end{equation}
Thus, if a D5-brane describing a Wilson line in the antisymmetric representation \( \text{A}_k \) sits at angle \( \theta_k \), then the D5-brane corresponding to the conjugate representation \( \overline{\text{A}_k} = \text{A}_{N-k} \) is located at \( \pi - \theta_k \), the antipodal   $S^4$ in $S^5$.

When comparing the F1-string charge of the initial and final D5-brane configurations in \eqref{chargeCC}, there is a subtlety: one must compute the charge in the  same gauge  for \( C_4 \). The antipodal map sends the brane to a region where a gauge regular at \( \theta = \pi \) is more appropriate, yielding a charge \( k_{\text{S}} \) given by 
\begin{equation}
k_{\text{S}} = \frac{N}{\pi} \left( \theta_{k_{\text{S}}} - \pi - \frac{1}{2} \sin 2\theta_{k_{\text{S}}} \right),
\end{equation}
with \( -N \leq k_{\text{S}} \leq 0 \). The charges in the two gauges are simply related by
\begin{equation}
  k_{\text{S}} =    k-N\,.
\end{equation}
This is the shift that turns \( -k \) into \( N - k \) in the final step of equation~\eqref{chargeCC}. Such shifts will occur whenever the initial and final five-branes (D5 or NS5) are related by an antipodal map -- i.e., they are located at \( \theta_k \) and \( \pi - \theta_k \), respectively. In conclusion, the action of the combined transformation \( \Omega (-1)^{F_L} R_{456789} \) in~\eqref{chargeCC} mirrors the effect of the charge conjugation operator \( \mathsf{C} \) on the \( k \)-th antisymmetric Wilson lines, as described in equation~\eqref{Cwilson}:
\begin{equation}
\mathsf{C}(W_{\text{A}_k}) = W_{\text{A}_{N - k}}.
\end{equation}

 While $\Omega(-1)^{F_L}$ preserves the D3-brane, it acts by charge conjugation on the $U(1)$ gauge field on the D3-brane worldvolume.\footnote{\label{myfootnote} More generally,  an $SL(2,\mathbb Z)$ duality  element  $M$ of Type IIB string theory induces an $SL(2,\mathbb Z)$ action on the $U(1)$ worldvolume gauge field strength  and its dual on the worldvolume of the D3-brane:
 \beq
 \begin{pmatrix}
 f\\
 *f
 \end{pmatrix}
\rightarrow M^{-T}
  \begin{pmatrix}
 f\\
 *f
 \end{pmatrix}\,.
 \label{actSL2D3}
 \eeq
 This can be inferred from the couplings $\int C_2\wedge f+ \int B_2\wedge *f$ of the D3-brane  and
 the $SL(2,\mathbb Z)$ action 
 $ \begin{pmatrix}
 C_2\\
 B_2
 \end{pmatrix}
 \rightarrow M 
  \begin{pmatrix}
 C_2\\
 B_2
 \end{pmatrix}\,.$
 This  extends to $GL(2,\mathbb Z)$ transformations   $(-1)^{F_L}$ and $\Omega$,
 which act as $(-1)^{F_L}: (f,*f)\rightarrow (-f,*f)$ and $\Omega: (f,*f)\rightarrow (f,-*f)$.} 
 Therefore, the transformation \( \Omega(-1)^{F_L} \) flips the sign of the background electric field on the D3-brane, resulting in a configuration we denote by D3\(_{-k}\). This realizes, on the D3-brane worldvolume, the map \( \text{F1} \rightarrow \overline{\text{F1}} \) induced by \( \Omega(-1)^{F_L} \).
While a D3\(_{k} \) located at \( \theta = 0 \) describes a Wilson loop in the \( k \)-th symmetric representation, a D3\(_{-k} \) located at \( \theta = \pi \) describes a Wilson loop in the complex conjugate representation. We have thus  reproduced in \eqref{chargeCC} the  action of charge conjugation  in equation~\eqref{Cwilson}:
\begin{equation}
\mathsf{C}(W_{\text{S}_k}) = W_{\overline{\text{S}_k}}.
\end{equation}
 
 The action of $\Omega(-1)^{F_L}R_{456789}$ on the bulk description of 't Hooft lines in representations $\text{A}_k$ and $\text{S}_k$ is:
   \beq
 \begin{aligned}
  \text{NS5}_k~   AdS_2\times S^4~ \text{at}~\theta_k &\rightarrow \overline{\text{NS5}}_{-k}~  AdS_2\times \overline{S^4} ~ \text{at}~\pi-\theta_k=
  \text{NS5}_{N-k}~   AdS_2\times S^4 ~ \text{at}~\pi-\theta_k \\  
  \text{D3}_k~ AdS_2\times S^2 ~\text{at}~\theta=0  &\rightarrow
  \text{D3}_{-k} ~ AdS_2\times S^2~\text{at}~\theta=\pi\,.
  \label{chargeCCC}
\end{aligned}
 \eeq
 There are two differences with respect to our analysis in \eqref{chargeCC}. First the electric field on NS5-brane worldvolume induces D1-string charge, not F1-string charge. The D3-brane now carries magnetic flux over $S^2$, inducing D1-string charge. Since D1-string charge is also odd under $\Omega (-1)^{F_L}$ this explains the sign flip in the subscript. Everything else goes verbatim the discussion above, and we have reproduced  from the bulk description the action of charge conjugation on 't Hooft loops~\eqref{conT}:
\begin{equation}
\begin{aligned}
\mathsf{C}(T_{\text{A}_k}) &= T_{\text{A}_{N - k}}\\
\mathsf{C}(T_{\text{S}_k}) &= T_{\overline{\text{S}_{k}}}\,.
\end{aligned}
\end{equation}

\begin{figure}[H]
    \centering
    \includegraphics[width=0.45\textwidth]{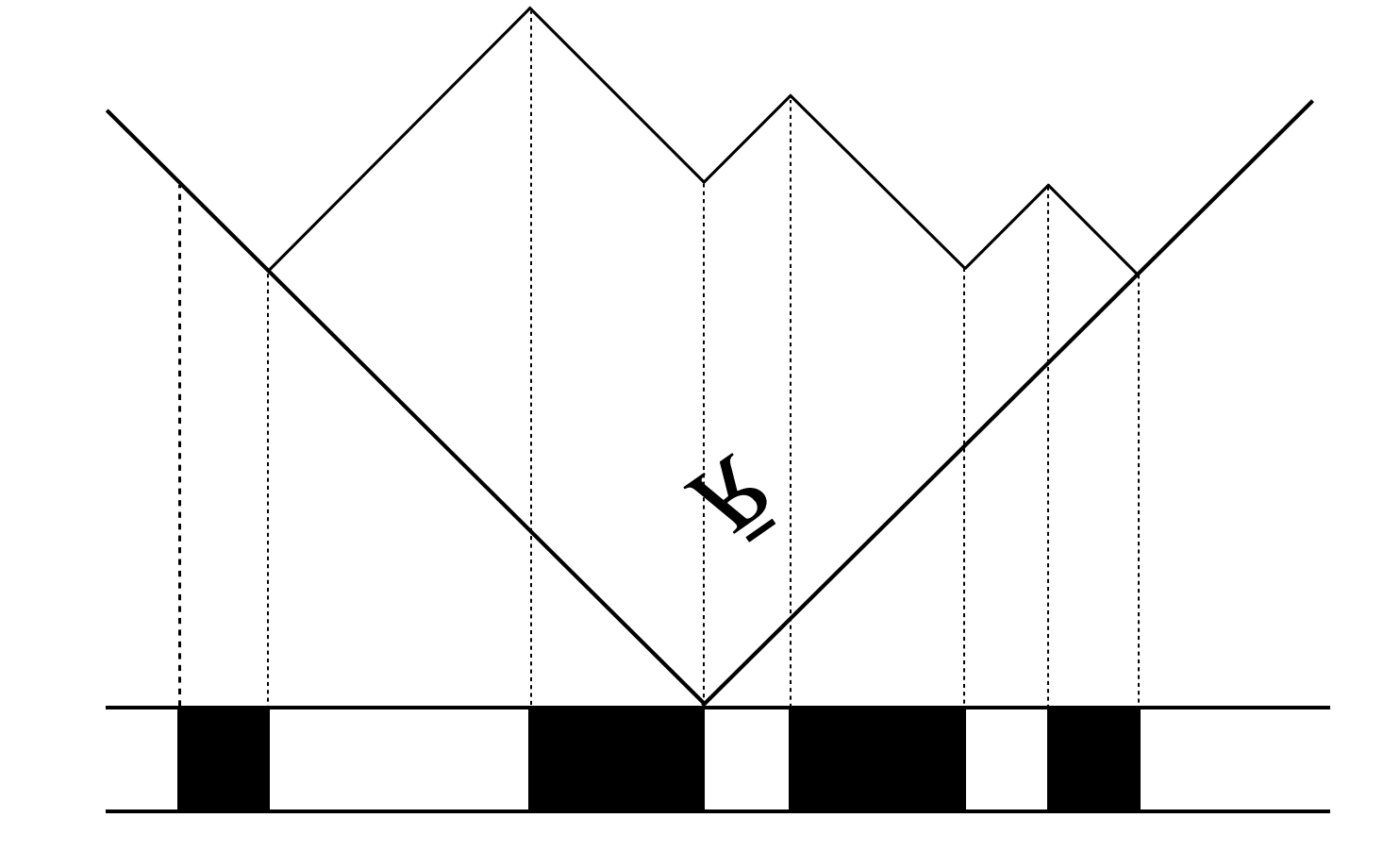}
    \caption{The coloring obtained by reflecting the \(x\)-axis of the configuration in Figure~\ref{fig:bubblingw}, 
as induced by the action of \(\Omega(-1)^{F_L} R_{456789}\), 
corresponds to the boundary condition describing a Wilson line in the complex conjugate representation \(\overline  {R}\).}
    \label{fig:bubblingwc}
  \end{figure}

 We turn to a     Wilson loop in a representation $R$. How does $\Omega(-1)^{F_L} R_{456789}$ act on the boundary data  determining the Type IIB bubbling solution    dual to a Wilson loop in a representation $R$? Let \((x, y)\), with \(y \geq 0\), be the coordinates on the upper half-plane over which a dual  asymptotically  \(AdS_5 \times S^5\) geometry is constructed by fibering \(AdS_2 \times S^2 \times S^4\). Then, the action of    \(\Omega\, (-1)^{F_L} R_{456789}\) on these coordinates is given by:\footnote{\label{footnoteW} This is inferred by 
 noting  that $x=\cosh \rho\cos\theta$ and $y=\sinh\rho \sin\theta$ for the $AdS_5\times S^5$ vacuum solution
 \[
 \cosh^2 \rho\, ds_{AdS_2}^2+d\rho^2+\sinh^2 \rho\, d\Omega_2+d\theta^2+\sin^2\theta\, d\Omega_4\,.
 \] The antipodal map $R_{456789}$ includes $\theta\rightarrow \pi-\theta$ so that 
 $(x,y)\rightarrow (-x,y)$ under  \(\Omega\, (-1)^{F_L} R_{456789}\).}
 \beq
 \begin{aligned}
  x&\rightarrow - x\\
   y&\rightarrow y\,.
 \end{aligned}
 \eeq
Of course, it also acts on the fluxes     and on the $S^4$ by the antipodal map. Therefore,  $\Omega(-1)^{F_L} R_{456789}$ flips the coloring along the $x$-axis at $y=0$. Pleasingly, the mirror coloring (up to translation)~\cite{Yamaguchi:2006te} is the coloring of the complex conjugate representation $\overline{R}$  (see figure \ref{fig:bubblingwc} and compare   with  figure \ref{fig:bubblingw}). Therefore, the action of $\Omega(-1)^{F_L} R_{456789}$  on the geometry dual to a Wilson line in a representation $R$ realizes the action of charge conjugation
 \eqref{CwilsonR} in ${\cal N}=4$ SYM:
 \beq
 {\mathsf C}(W_{R}) = W_{  \overline { {R}}}\,.
 \eeq
  \noindent
An analogous analysis applies to the bulk description of 't Hooft lines and proceeds in parallel with 
the discussion for Wilson lines. As expected, it correctly reproduces the action of charge conjugation 
on  't Hooft lines, as given in \eqref{conTR}.

A surface operator with   Levi group $\mathfrak{su}(N-1)\times \mathfrak{u}(1)$ is described by a D3-brane  with an $AdS_3 \times S^1$ embedding \eqref{surfaceD3}. The antipodal map $R_{456789}$ on $S^5$ acts on the coordinates of $S^5$ in \eqref{S5met} as
\beq
\vartheta\rightarrow \vartheta\qquad \phi\rightarrow \pi+\phi\,,
\label{actantipp}
\eeq 
as well as the antipodal map on $S^3$. Also, \(\Omega(-1)^{F_L}\)  acts as charge conjugation on the D3-brane worldvolume gauge field. This, combined with the parameter mapping \eqref{surfacemap},\footnote{$(\beta,\gamma)$ are reversed by \eqref{actantipp} and $(\alpha,\eta)$ are reversed by charge conjugation on the D3-brane worldvolume.}  shows that $\Omega(-1)^{F_L} R_{456789}$ acting on the D3-brane:
\beq
 \text{D3}_{(\alpha,\beta,\gamma,\eta)} ~ \text{on}~  AdS_3 \times S^1  \rightarrow  \text{D3}_{(-\alpha,-\beta,-\gamma,-\eta)}~ \text{on}~ AdS_3 \times S^1\,,
\eeq
realizes  the action of $\mathsf C$ on the surface operator \eqref{actsurf}:
 \beq
 \mathsf C\left({\cal O}_{\Sigma}(\alpha,\beta,\gamma,\eta)  \right)={\cal O}_{\Sigma}(-\alpha,-\beta,-\gamma,-\eta)\,.
 \label{actsurfa}
 \eeq
 
 For a general surface operator described by a bubbling geometry, the action of \( R_{456789} \) on the half-space \( \mathbb{R}^3_+ \), with coordinates \( (\vec{x}, y) \) and \( y \geq 0 \), over which the asymptotically \( AdS_5 \times S^5 \) solution is constructed by fibering \( AdS_3 \times S^1 \times S^3 \), is given by:\footnote{This is deduced by 
 noting  that in the $AdS_5\times S^5$ vacuum solution \eqref{AdS}\eqref{S5met}: 
 \begin{align*}
x^1 + i x^2 &= r\, e^{i(\psi + \phi)}\,, \\ 
r &= \sinh u\, \cos\vartheta\,, \\ 
y &= \cosh u\, \sin\vartheta\,, \\ 
\chi &= \frac{1}{2}(\psi - \phi)\,,
\end{align*}
where $\chi$ is the $S^1$ fiber.
The antipodal map $R_{456789}$  \eqref{actantipp}  implies that  
 $(\vec x,y)\rightarrow (-\vec x,y)$.}
 \beq
 \begin{aligned}
  \vec x&\rightarrow - \vec x\\
   y&\rightarrow y\,.
 \end{aligned}
 \eeq
Given that  \(\Omega(-1)^{F_L}\) flips the sign of $B_2$ and $C_2$, and the mapping in \eqref{mapsurfb},
$\Omega(-1)^{F_L} R_{456789}$  acting on the surface operator bubbling geometry realizes the action of $\mathsf C$ for an arbitrary surface 
operator~\eqref{actsurfa}.

\subsection{Holographic ${\mathsf P}$-symmetry}

The parity symmetry  \eqref{p1act} of \({\cal N}=4\) SYM is realized in Type IIB string theory  in $AdS_5\times S^5$ by: 
 \beq
 \mathsf P\longleftrightarrow  (-1)^{F_L}  R_{3456}\,.
 \eeq
 The reflection \( R_3 \) acts geometrically on \( AdS_5 \times S^5 \) by reversing the \( x^3 \subset \mathbb{R}^{1,3} \subset AdS_5 \) coordinate, while \( R_{456} \) flips half of the embedding coordinates of the \( S^5 \). Although the five-form flux supporting the \( AdS_5 \times S^5 \) geometry is odd under \( (-1)^{F_L} \), when combined with \( R_{3456} \), the flux remains invariant. Moreover, the reflection \( R_{3456} \) preserves the orientation of spacetime, and    \(   (-1)^{F_L} R_{3456} \) acts on the axio-dilaton as \( \tau \rightarrow -\bar{\tau} \), in the same way as  \( \mathsf{P} \) acts on the exactly marginal coupling~\eqref{Pontau}.  

The symmetry \( (-1)^{F_L} R_{3456} \) acts trivially on both the metric and the RR four-form \( C_4 \), despite \( C_4 \) being odd under \( (-1)^{F_L} \). Consequently, the nontrivial action of \( (-1)^{F_L} R_{3456} \) on the supergravity mode dual to the chiral primary operator \( \mathcal{O}_\Delta \) arises from its action on the spherical harmonic on \( S^5 \). This reproduces the action of \( \mathsf{P} \)  on chiral primary operators as described in equation~\eqref{parityCPO}.

 In section \ref{sec:P}, we classified line  operators according to whether they couple to a scalar field that is reflected under the action of \( \mathsf{P} \) in~\eqref{p1act}. The brane dual to an operator coupling to a scalar $\phi^{\dot A}$ that is not reflected    remains fixed under the action of \( R_{456} \), and thus stays at the same point \( P \subset S^5 \). This implies that if the dual brane wraps an \( S^4 \), its orientation is  reversed, yielding \( \overline{S^4} \), since \( R_{456} \) reflects an odd number of embedding coordinates.
Conversely, the brane dual to an operator coupling to a scalar $\phi^{A}$ that  is reflected,  is mapped under \( R_{456} \) to the antipodal point on \( S^5 \), which we denote by \( \overline{P} \). In this case, if the dual brane wraps $S^4$,  its orientation is preserved, as an even number coordinates are inverted. 
We consider  operators that are either in the $x^3=0$ fixed locus of  \( \mathsf{P} \) or extended along the $x^3$ direction, distinguishing the latter by a hat. The $AdS_2$ geometry of a brane dual to a hatted operator    has its orientation reversed under 
 \( R_{3456} \) and maps to $\overline{AdS_2}$. Conversely, if the   brane dual to      an unhatted operator wraps $S^2$,    its orientation is reversed under 
 \( R_{3456} \) and maps to $\overline{S^2}$. For a summary   the action of  \( R_{3456} \)    depending  on   line operator considered, see table \ref{tab:parity-map}.
\begin{table}[ht]
\centering
\renewcommand{\arraystretch}{1.4} 
\begin{tabular}{|>{\centering\arraybackslash}m{2.2cm}|
                  >{\centering\arraybackslash}m{4.2cm}|
                  >{\centering\arraybackslash}m{4.2cm}|}
\hline
\(\phi^{A}\)         & \(P \longrightarrow \overline{P}\)                & \(S^4 \longrightarrow S^4\) \\
\(\phi^{\dot{A}}\)   & \(P \longrightarrow P\)                           & \(S^4 \longrightarrow \overline{S}^4\) \\
\hline\hline
\rule{0pt}{2.8ex}\(\mathcal{L}\)      & \(\mathrm{AdS}_2 \longrightarrow \mathrm{AdS}_2\) & \(S^2 \longrightarrow \overline{S}^2\) \\
\(\hat{\mathcal{L}}\)& \(\mathrm{AdS}_2 \longrightarrow \overline{\mathrm{AdS}}_2\) & \(S^2 \longrightarrow S^2\) \\
\hline
\end{tabular}
\caption{Action of $R_{3456}$ on  $S^5$ location $P$, orientation of $AdS_2, S^2$ and $S^4$ according to     scalar coupling of line operator and choice of line ${\cal L}$ or \(\hat{\mathcal{L}}\). $\phi^{A}$ is odd and $\phi^{\dot A}$ even under  $R_{456}$.  }
\label{tab:parity-map}
\end{table}

The action of $(-1)^{F_L}  R_{3456}$ on the  branes dual to Wilson and 't Hooft line operators in the fundamental representation is:\footnote{Since $\mathsf{P}$ preserves a static Wilson line and maps a static 't~Hooft line to its conjugate, the corresponding worldsheet symmetry must act as ${\rm F1}\to{\rm F1}$ and ${\rm D1}\to\overline{{\rm D1}}$. This action is implemented by $(-1)^{F_L}$.}
\beq
\begin{aligned}
W_{\text{fund}}(\phi^{\dot A}):~  \text{F1 on}~ AdS_2~\text{at}~\text{P}&\rightarrow {\text F1}~ AdS_2~\text{at}~\text{P}\,, \\ 
T_{\text{fund}}(\phi^{A}):~  \text{D1 on}~ AdS_2~\text{at}~\text{P}&\rightarrow {\overline {\text D1}}~ AdS_2~\text{at}~{\overline{\text{P}}}\,, \\ 
\hat W_{\text{fund}}(\phi^{ A}):~  \text{F1 on}~ AdS_2~\text{at}~\text{P}&\rightarrow   {\text F1}~ \text{on}~ \overline{AdS_2}~\text{at}~{\overline{\text{P}}}={\overline {\text F1}}~ \text{on}~ AdS_2~\text{at}~{\overline{\text{P}}}\,, \\ 
\hat T_{\text{fund}}(\phi^{ \dot A}):~  \text{D1 on}~ AdS_2~\text{at}~\text{P}&\rightarrow  {\overline {\text D1}}~ \text{on}~ \overline{AdS_2}~\text{at}~\text{P}= {\text D1}~ \text{on}~ AdS_2~\text{at}~{\text{P}}\,.
\end{aligned}
\eeq
For lines in the $k$-th symmetric representation:
\beq
\begin{aligned}
W_{\text{S}_k}(\phi^{\dot A}):~& \text{D3}_k~ \text{on}~ AdS_2\times S^2~  \text{at}~  \text{P} \rightarrow
\overline{\text{D3}}_k~ \text{on}~ AdS_2\times \overline{S^2}~  \text{at}~  \text{P}=\text{D3}_k~ \text{on}~ AdS_2\times S^2~  \text{at}~  \text{P}\,, 
 \\  
T_{\text{S}_k}(\phi^{A}):~& \text{D3}_k~ \text{on}~ AdS_2\times S^2~  \text{at}~  \text{P} \rightarrow
\overline{\text{D3}}_{-k}~ \text{on}~ AdS_2\times \overline{S^2}~  \text{at}~  \overline{\text{P}}=\text{D3}_{-k}~ \text{on}~ AdS_2\times S^2~  \text{at}~ \overline{\text{P}}\,, 
 \\  
   \hat W_{\text{S}_k}(\phi^{A}):~& \text{D3}_k~ \text{on}~ AdS_2\times S^2~  \text{at}~  \text{P} \rightarrow
\overline{\text{D3}}_{-k}~ \text{on}~  \overline{AdS_2}\times  S^2 ~  \text{at}~  \overline{\text{P}}=\text{D3}_{-k}~ \text{on}~ AdS_2\times S^2~  \text{at}~  \overline{\text{P}}\,, \\
\hat T_{\text{S}_k}(\phi^{\dot A}):~& \text{D3}_k~ \text{on}~ AdS_2\times S^2~  \text{at}~  \text{P} \rightarrow
\overline{\text{D3}}_{k}~ \text{on}~  \overline{AdS_2}\times S^2~  \text{at}~  \text{P}=\text{D3}_{k}~ \text{on}~ AdS_2\times S^2~  \text{at}~   \text{P} \,,
\end{aligned}
\eeq
while for lines in the $k$-th antisymmetric representation:
\beq
\begin{aligned}
W_{\text{A}_k}(\phi^{\dot A}):~& \text{D5}_k~ \text{on}~ AdS_2\times S^4~  \text{at}~  \theta_k \rightarrow
\overline{\text{D5}}_k~ \text{on}~ AdS_2\times \overline{S^4}~  \text{at}~  \theta_k=\text{D5}_k~ \text{on}~ AdS_2\times S^4~  \text{at}~  \theta_k\,,\\ 
T_{\text{A}_k}(\phi^{A}):~& \text{NS5}_k~ \text{on}~ AdS_2\times S^4~  \text{at}~  \theta_k \rightarrow
\text{NS5}_{N-k}~ \text{on}~ AdS_2\times S^4 ~  \text{at}~  \pi-\theta_k\,,\\ 
\hat W_{\text{A}_k}(\phi^{A}):~& \text{D5}_k~ \text{on}~ AdS_2\times S^4~  \text{at}~  \theta_k \rightarrow
\overline{\text{D5}}_{-k}~ \text{on}~ \overline{AdS_2}\times  S^4 ~  \text{at}~  \pi-\theta_k=\text{D5}_{N-k}~ \text{on}~ AdS_2\times S^4~  \text{at}~ \pi- \theta_k\,,\\ 
\hat T_{\text{A}_k}(\phi^{\dot A}):~& \text{NS5}_k~ \text{on}~ AdS_2\times S^4~  \text{at}~  \theta_k \rightarrow
 \text{NS5}_k~ \text{on}~ \overline{AdS_2}\times  \overline{S^4} ~  \text{at}~   \theta_k=\text{NS5}_k~ \text{on}~ AdS_2\times S^4~  \text{at}~  \theta_k\,.
\end{aligned}
\eeq
Using the dictionary between branes and line operators in \eqref{Wilsonmap} and \eqref{hooftmap}, we determine the transformation of the bulk description of a given line operator under ${\mathsf P}$. This analysis reveals that the combination $(-1)^{F_L} R_{3456}$ precisely mirrors the action of ${\mathsf P}$ on the line operators in ${\cal N}=4$ SYM in table \ref{linePCP}:
\beq
\begin{aligned}
\mathsf{P}(W_{R}(\phi^{\dot A})) &= W_{R}(\phi^{\dot A}) \,, \\
\mathsf{P}(T_{R}(\phi^{A})) &= T_{\overline{R}}(\phi^{A}) \,, \\
\mathsf{P}(\hat W_{R}(\phi^{A})) &= \hat W_{\overline{R}}(\phi^{A})\,, \\
\mathsf{P}(\hat{T}_{R}(\phi^{\dot{A}})) &= \hat{T}_{R}(\phi^{\dot{A}}) \,.
\label{linePCPa}
\end{aligned}
\eeq
For a general representation, the action of \( (-1)^{F_L} R_{3456} \) on the coloring   of the $x$-axis defining the dual supergravity solution depends on whether the line operator couples to a scalar \( \phi^A \) that is reflected under \( \mathsf{P} \), or to a scalar \( \phi^{\dot A} \) that is not. In the former case, \( (-1)^{F_L} R_{3456} \) reflects the colored \( x \)-axis,\ while in the latter it does not,\footnote{In the former case, the action of \( R_{456} \) includes the reflection \( \theta \to \pi - \theta \); in the latter, it does not (cf. footnote
\ref{footnoteW}).
} thereby reproducing equation~\eqref{linePCPa}.

 We now consider the action of ${\mathsf  P}$ on surface operators. They are supported on the two-dimensional plane $\Sigma = \mathbb{R}^2$, with local coordinates $(x^2, x^3)$. The surface operator couples to a complex scalar field as described in \eqref{surface1}, where the scalar can belong either to the set $\phi^A$, which is reflected under ${\mathsf  P}$, or to the set $\phi^{\dot A}$, which remains invariant.

Using  that $(-1)^{F_L}$ maps a  D3 to a $\overline{D3}$, that   $\eta=\oint {\tilde a\over 2\pi}$  and $\tilde a$ is odd under $(-1)^{F_L}$ (see footnote \ref{myfootnote}),  and that $R_3$ reverses the orientation and maps $AdS_3$ to $\overline{AdS_3}$, we find the
action of  $(-1)^{F_L} R_{3456}$   on the D3-branes dual to surface operators:
 \beq
\begin{aligned}
 {\cal O}_{\Sigma}(\alpha,\beta,\gamma,\eta) [\phi^{\dot A}]:~   \text{D3}_{(\alpha,\beta,\gamma,\eta)} ~ \text{on}~  AdS_3 \times S^1 &\rightarrow \overline{\text{D3}}_{(\alpha,\beta,\gamma,-\eta)}~ \text{on}~ \overline{AdS_3}\times S^1\,,\\
  &=\text{D3}_{(\alpha,\beta,\gamma,-\eta)} ~ \text{on}~  AdS_3 \times S^1\,, \\[+2pt]
 {\cal O}_{\Sigma}(\alpha,\beta,\gamma,\eta) [\phi^{A}]:~   \text{D3}_{(\alpha,\beta,\gamma,\eta)} ~ \text{on}~  AdS_3 \times S^1 & \rightarrow \overline{\text{D3}}_{(\alpha,-\beta,-\gamma,-\eta)}~ \text{on}~ \overline{AdS_3}\times  S^1\,, \\
& =\text{D3}_{(\alpha,-\beta,-\gamma,-\eta)} ~ \text{on}~  AdS_3 \times S^1\,.
\end{aligned}
\label{surfaceD3aaa}
\eeq
This reproduces, from a bulk perspective, the action of $\mathsf P$ symmetry of ${\cal N}=4$ SYM on surface operators~\eqref{actonsurfpp}:
\begin{equation}
\begin{aligned}
\mathsf{P}\bigl(\mathcal{O}_\Sigma(\alpha,\beta,\gamma,\eta)[\phi^{\dot A}]\bigr)
  &= \mathcal{O}_\Sigma(\alpha,\beta,\gamma,-\eta)[\phi^{\dot A}], \\[6pt]
\mathsf{P}\bigl(\mathcal{O}_\Sigma(\alpha,\beta,\gamma,\eta)[\phi^{A}]\bigr)
  &= \mathcal{O}_\Sigma(\alpha,-\beta,-\gamma,-\eta)[\phi^{A}]\,.
  \label{adtsurfpppp}
\end{aligned}
\end{equation}
For a general surface operator, \( (-1)^{F_L} R_{3456} \) inverts   the location of  the point sources $\vec x_l\rightarrow -\vec x_l$ specifying the dual supergravity solution if the surface operator couples to a complex scalar reflected by $\mathsf P$, and it doesn't otherwise. This combined with the fact that $(-1)^{F_L}$ flips $C_2$ but not $B_2$,
and the parameter mapping \eqref{mapsurfb}, we  realize the action of ${\cal P}$ on surface operators \eqref{adtsurfpppp} as \( (-1)^{F_L} R_{3456} \) on the dual bubbling supergravity solutions.

\subsection{Holographic ${\mathsf T}$-symmetry}

The time-reversal symmetry  \eqref{Tact} in \({\cal N}=4\) SYM is realized in Type IIB string theory  in $AdS_5\times S^5$ by: 
 \beq
 \mathsf T\longleftrightarrow  \Omega R_{0789}\,.
 \eeq
 The reflection \( R_0 \) acts geometrically on \( AdS_5 \times S^5 \) by reversing the \( x^0 \subset \mathbb{R}^{1,3} \subset AdS_5 \) coordinate, while \( R_{789} \) flips half of the embedding coordinates of the \( S^5 \). Although the five-form flux supporting the \( AdS_5 \times S^5 \) geometry is odd under \( \Omega \), when combined with \( R_{0789} \), the flux remains invariant. Moreover, the reflection \( R_{0789} \) preserves the orientation of spacetime, and    \(   \Omega R_{0789} \) acts on the axio-dilaton as \( \tau \rightarrow -\bar{\tau} \), in the same way as  \( \mathsf{T} \) acts on the exactly marginal coupling~\eqref{Tontau}.

The symmetry \( \Omega R_{0789}\) acts trivially on both the metric and the RR four-form \( C_4 \), despite \( C_4 \) being odd under \( \Omega \). Consequently, the nontrivial action of \( \Omega R_{0789}\)  on the supergravity mode dual to the chiral primary operator \( \mathcal{O}_\Delta \) arises from its action on the spherical harmonic on \( S^5 \). This reproduces the action of \( \mathsf T \) on chiral primary operators as described in equation~\eqref{TonCPO}.

 In section \ref{sec:T}, we classified line operators according to whether they couple to a scalar field that is reflected under the action of \( \mathsf{T} \), as defined in~\eqref{Tact}. $\phi^{\dot A}$ and $\phi^A$
 are odd and even under \( \mathsf{T} \) respectively. The brane dual to an operator coupling to a scalar that is not reflected  remains fixed under the action of \( R_{789} \), and thus stays at the same point \( P \subset S^5 \). This, in turn, implies that if the dual brane wraps $S^4$, that its orientation is  reserved and becomes $\overline{S^4}$.
Conversely, the brane dual to an operator coupling to a scalar that  is reflected,  is mapped under \( R_{789} \) to the antipodal point on \( S^5 \), which we denote by \( \overline{P} \). In this case, if the dual brane wraps $S^4$,  its orientation is preversed.
Relatedly, we considered operators that are in the $x^0=0$ fixed locus of  \( \mathsf{T} \) or extended along the $x^0$ direction, distinguishing the latter by a hat. Therefore, the $AdS_2$ geometry brane dual of the hatted operators   has its orientation reversed under 
 \( R_{0789} \) and maps to $\overline{AdS_2}$. Conversely, if the dual brane   of an unhatted operator wraps $S^2$,    its orientation is reversed under 
 \( R_{0789} \) and maps to $\overline{S^2}$. 
 For a summary of how the action of  \( R_{0789} \)  depends on the choice of  line operator considered, see table \ref{tab:T-map}.

 \begin{table}[ht]
\centering
\renewcommand{\arraystretch}{1.4} 
\begin{tabular}{|>{\centering\arraybackslash}m{2.2cm}|
                  >{\centering\arraybackslash}m{4.2cm}|
                  >{\centering\arraybackslash}m{4.2cm}|}
\hline
\(\phi^{\dot A}\)         & \(P \longrightarrow \overline{P}\)                & \(S^4 \longrightarrow S^4\) \\
\(\phi^{ A}\)   & \(P \longrightarrow P\)                           & \(S^4 \longrightarrow \overline{S}^4\) \\
\hline\hline
\rule{0pt}{2.8ex}\(\mathcal{L}\)      & \(\mathrm{AdS}_2 \longrightarrow \mathrm{AdS}_2\) & \(S^2 \longrightarrow \overline{S}^2\) \\
\(\hat{\mathcal{L}}\)& \(\mathrm{AdS}_2 \longrightarrow \overline{\mathrm{AdS}}_2\) & \(S^2 \longrightarrow S^2\) \\
\hline
\end{tabular}
\caption{Action of $R_{0789}$ on  $S^5$ location $P$, orientation of $AdS_2, S^2$ and $S^4$ according to choice of scalar coupling and choice of line ${\cal L}$ or \(\hat{\mathcal{L}}\).}
\label{tab:T-map}
\end{table}

The action of $\Omega R_{0789}$ on the  branes dual to Wilson and 't Hooft line operators in the fundamental representation is:\footnote{Since $\mathsf{T}$ preserves a static Wilson line, maps a static 't~Hooft line to its conjugate and maps $AdS_2\rightarrow \overline{AdS_2}$, the corresponding worldsheet symmetry must act as ${\rm F1}\to\overline{\rm F1}$ and ${\rm D1}\to {{\rm D1}}$. This action is implemented by $\Omega$.}
 \beq
\begin{aligned}
W_{\text{fund}}(\phi^{\dot A}):~  \text{F1 on}~ AdS_2~\text{at}~\text{P}&\rightarrow   \overline{{\text F1}}~ AdS_2~\text{at}~{\overline{\text{P}}} \\ 
T_{\text{fund}}(\phi^{A}):~  \text{D1 on}~ AdS_2~\text{at}~\text{P}&\rightarrow \text D1 ~ AdS_2~\text{at}~\text{P} \\ 
\hat W_{\text{fund}}(\phi^{ A}):~  \text{F1 on}~ AdS_2~\text{at}~\text{P}&\rightarrow    \overline{{\text F1}}~ \text{on}~ \overline{AdS_2}~\text{at}~ \text{P}= \text{F1}~ \text{on}~ AdS_2~\text{at}~\text{P} \\ 
\hat T_{\text{fund}}(\phi^{ \dot A}):~  \text{D1 on}~ AdS_2~\text{at}~\text{P}&\rightarrow   {\text D1} ~ \text{on}~ \overline{AdS_2}~\text{at}~\overline{\text{P}}= \overline{{\text D1}}~ \text{on}~ AdS_2~\text{at}~\overline{{\text{P}}}\,.
\end{aligned}
\eeq 
 For lines in the $k$-th symmetric representation:
\beq
\begin{aligned}
W_{\text{S}_k}(\phi^{\dot A}):~& \text{D3}_{k}~ \text{on}~ AdS_2\times S^2~  \text{at}~  \text{P} \rightarrow
 \overline{\text{D3}}_{-k}~ \text{on}~ AdS_2\times \overline{S^2}~  \text{at}~  \overline{\text{P}}=  \text{D3}_{-k}~ \text{on}~ AdS_2\times S^2~  \text{at}~  \overline{\text{P}} 
 \\  
T_{\text{S}_k}(\phi^{A}):~& \text{D3}_k~ \text{on}~ AdS_2\times S^2~  \text{at}~  \text{P} \rightarrow
 \overline{\text{D3}}_{k}~ \text{on}~ AdS_2\times \overline{S^2}~  \text{at}~   \text{P} =\text{D3}_{k}~ \text{on}~ AdS_2\times S^2~  \text{at}~  \text{P}
 \\  
   \hat W_{\text{S}_k}(\phi^{A}):~& \text{D3}_k~ \text{on}~ AdS_2\times S^2~  \text{at}~  \text{P} \rightarrow
 \overline{\text{D3}}_{k}~ \text{on}~   \overline{AdS_2} \times  S^2 ~  \text{at}~   \text{P} =\text{D3}_{k}~ \text{on}~ AdS_2\times S^2~  \text{at}~   \text{P}  \\
\hat T_{\text{S}_k}(\phi^{\dot A}):~& \text{D3}_k~ \text{on}~ AdS_2\times S^2~  \text{at}~  \text{P} \rightarrow
\overline{\text{D3}}_{-k}~ \text{on}~  \overline{AdS_2}\times S^2~  \text{at}~  \overline{\text{P}}=  \text{D3}_{-k}~ \text{on}~ AdS_2\times S^2~  \text{at}~   \overline{\text{P}} \,,
\end{aligned}
\eeq
while for lines in the $k$-th antisymmetric representation:
\beq
\begin{aligned}
W_{\text{A}_k}(\phi^{\dot A}):~& \text{D5}_k~ \text{on}~ AdS_2\times S^4~  \text{at}~  \theta_k \rightarrow
 \text{D5}_{-k}~ \text{on}~ AdS_2\times  S^4 ~  \text{at}~ \pi- \theta_k=\text{D5}_{N-k}~ \text{on}~ AdS_2\times S^4~  \text{at}~ \pi- \theta_k\\ 
T_{\text{A}_k}(\phi^{A}):~& \text{NS5}_k~ \text{on}~ AdS_2\times S^4~  \text{at}~  \theta_k \rightarrow
\overline{\text{NS5}}_{k}~ \text{on}~ AdS_2\times \overline{S^4} ~  \text{at}~  \theta_k
= \text{NS5} _{k}~ \text{on}~ AdS_2\times  S^4  ~  \text{at}~  \theta_k
\\ 
\hat W_{\text{A}_k}(\phi^{A}):~& \text{D5}_k~ \text{on}~ AdS_2\times S^4~  \text{at}~  \theta_k \rightarrow
 \text{D5}_{k}~ \text{on}~ \overline{AdS_2}\times  \overline{S^4} ~  \text{at}~ \theta_k
 =\text{D5}_k~ \text{on}~ AdS_2\times S^4~  \text{at}~  \theta_k\\
\hat T_{\text{A}_k}(\phi^{\dot A}):~& \text{NS5}_k~ \text{on}~ AdS_2\times S^4~  \text{at}~  \theta_k \rightarrow
 \overline{\text{NS5}}_{-k}~ \text{on}~ \overline{AdS_2}\times  S^4  ~  \text{at}~ \pi-  \theta_k=\text{NS5}_{N-k}~ \text{on}~ AdS_2\times S^4~  \text{at}~  \theta_k\,.
\end{aligned}
\eeq
 This analysis reveals that the combination $\Omega R_{0789}$ precisely mirrors the action of ${\mathsf T}$ on the line operators in ${\cal N}=4$ SYM in table \ref{WilsonTCT}:
\beq
\begin{aligned}
\mathsf{T}(W_{R}(\phi^{\dot A})) &= W_{\overline{R}}(\phi^{\dot A}) \,, \\
\mathsf{T}(T_{R}(\phi^{A})) &= T_{{R}}(\phi^{A}) \,, \\
\mathsf{T}(\hat W_{R}(\phi^{A})) &= \hat W_{R}(\phi^{A})\,, \\
\mathsf{T}(\hat{T}_{R}(\phi^{\dot{A}})) &= \hat{T}_{\overline{R}}(\phi^{\dot{A}}) \,.
\label{linePCPab}
\end{aligned}
\eeq
The same conclusion is reached from the action of $\Omega R_{0789}$ on the  dual bubbling geometry description of a line operator in a general representation $R$.

Using  that $\Omega$ maps a  D3 to a $\overline{D3}$, that    $\alpha=\oint { a\over 2\pi}$ and $a$ is odd under $\Omega$ (see footnote \ref{myfootnote}),  and that $R_0$ reverses the orientation and maps $AdS_3$ to $\overline{AdS_3}$, we find the
action of  $\Omega R_{0789}$   on the D3-branes dual to surface operators: 
 \beq
\begin{aligned}
 {\cal O}_{\Sigma}(\alpha,\beta,\gamma,\eta) [\phi^{ A}]:~   \text{D3}_{(\alpha,\beta,\gamma,\eta)} ~ \text{on}~  AdS_3 \times S^1 &\rightarrow \overline{\text{D3}}_{(-\alpha,\beta,\gamma,\eta)}~ \text{on}~ \overline{AdS_3}\times S^1\\
  &=\text{D3}_{(-\alpha,\beta,\gamma,\eta)} ~ \text{on}~  AdS_3 \times S^1 \\[+2pt]
 {\cal O}_{\Sigma}(\alpha,\beta,\gamma,\eta) [\phi^{\dot A}]:~   \text{D3}_{(\alpha,\beta,\gamma,\eta)} ~ \text{on}~  AdS_3 \times S^1 & \rightarrow \overline{\text{D3}}_{(-\alpha,-\beta,-\gamma,\eta)}~ \text{on}~ \overline{AdS_3}\times  S^1 \\
& =\text{D3}_{(-\alpha,-\beta,-\gamma,\eta)} ~ \text{on}~  AdS_3 \times S^1\,.
\end{aligned}
\label{surfaceD3aaa}
\eeq
This reproduces, from a bulk perspective, the action of $\mathsf T$ symmetry of ${\cal N}=4$ SYM on surface operators~\eqref{actonsurfttt}:
\begin{equation}
\begin{aligned}
\mathsf{T}\bigl(\mathcal{O}_\Sigma(\alpha,\beta,\gamma,\eta)[\phi^{ A}]\bigr)
  &= \mathcal{O}_\Sigma(-\alpha,\beta,\gamma,\eta)[\phi^{A}], \\[6pt]
\mathsf{T}\bigl(\mathcal{O}_\Sigma(\alpha,\beta,\gamma,\eta)[\phi^{\dot A}]\bigr)
  &= \mathcal{O}_\Sigma(-\alpha,-\beta,-\gamma,\eta)[\phi^{\dot A}]\,.
  \label{actonsurftttt}
\end{aligned}
\end{equation}
For a general surface operator, \( \Omega R_{0789} \) inverts   the location of  the point sources $\vec x_l\rightarrow -\vec x_l$ specifying the dual supergravity solution if the surface operator couples to a complex scalar reflected by $\mathsf T$, and it doesn't otherwise. This combined with the fact that $\Omega$ flips $B_2$ but not $C_2$,
and the parameter mapping \eqref{mapsurfb}, we  realize the action of ${\cal T}$ on surface operators \eqref{actonsurftttt} as \( \Omega R_{0789} \) on the dual bubbling supergravity solutions.

\subsection{Holographic $\mathsf {CPT}$-symmetry}

The $\mathsf{CPT}$ symmetry  \eqref{CPTact} of \({\cal N}=4\) SYM is realized in Type IIB string theory  in $AdS_5\times S^5$ by: 
 \beq
 \mathsf{CPT}\longleftrightarrow    R_{03}\,.
 \eeq
 The reflection  \( R_{03} \) acts geometrically on \( AdS_5 \times S^5 \) by reversing the \( (x^0,x^3) \subset \mathbb{R}^{1,3} \subset AdS_5 \) coordinates.  \( R_{03} \) is an      orientation preserving symmetry of the 
 \( AdS_5 \times S^5 \) background that acts on the axio-dilaton as \( \tau \rightarrow  {\tau} \), in the same way as  \( \mathsf{CPT} \) acts on the exactly marginal coupling~\eqref{CPTact}. 
 
 The branes and geometries  dual to Wilson and 't Hooft line operators  in the plane fixed by ${\cal P}{\cal T}$ are trivially acted
 by $R_{03}$, they are left invariant. This realizes the action of $\mathsf{CPT}$ on the line operators \eqref{CPTW1}
 \begin{equation}
\begin{aligned}
\mathsf{CPT}(W_{R})  &= W_{R} \,, \\
\mathsf{CPT}(T_{R})  &= T_{R} \,.
\end{aligned}
\end{equation}
The branes   dual to Wilson and 't Hooft lines along say $x^0$ or $x^3$ and not sourcing a scalar field, 
have their orientation reversed, and realize the expected action:
\begin{equation}
\begin{aligned}
\mathsf{CPT}(\hat W_{R})  &= \hat W_{\overline{R}} \,, \\
\mathsf{CPT}(\hat T_{R})  &= \hat T_{\overline{R}} \,.
\end{aligned}
\end{equation}

Likewise, the branes and geometries  dual to a surface operator  supported on   $\Sigma = \mathbb{R}^{1,1}$, with local coordinates $(x^0, x^3)$, coupled to an
arbitrary complex scalar field, is fixed by \( R_{03} \) and realize the action of  $\mathsf{CPT}$ on the surface operators \eqref{CPTS1}:
\begin{equation}
\begin{aligned}
\mathsf{CPT}\bigl(\mathcal{O}_\Sigma(\alpha,\beta,\gamma,\eta)\bigr)
   = \mathcal{O}_\Sigma(\alpha,\beta,\gamma,\eta)\,.
 \end{aligned}
\end{equation}

The $\mathsf{CPT}$ symmetry of ${\cal N}=4$ SYM maps to the $\mathsf{CPT}_{\textrm{bulk}}$ symmetry of string theory in \( AdS_5 \times S^5 \):
\beq
 \mathsf{CPT}\longleftrightarrow     \mathsf{CPT}_{\textrm{bulk}}\,.
 \eeq

\section{Charge Conjugation Monodromy Surface Operator}
\label{sec:surface}

 A quantum field theory with a discrete symmetry group admits   codimension-two monodromy defects, defined by imposing twisted boundary conditions on the fields $\varphi$ as they encircle the defect. For each symmetry element $g$, the corresponding monodromy defect $\mathsf{M}_g$ imposes the boundary condition
\beq
\mathsf{M}_g: \quad \varphi(z\, e^{2\pi i}) = g \cdot \varphi(z)\,,
\eeq
where $g \cdot \varphi$ denotes the action of  $g$ on $\varphi$, and $z$ is a complex coordinate in the plane transverse to the codimension-two defect.  The monodromy defect is supported at the boundary of the topological defect implementing the action of the symmetry $g$, the topological defect being represented by the ``branch cut" in Figure \ref{fig:branchcut}.
 \begin{figure}[H]
  \centering
  \begin{tikzpicture}

    \draw[dashed, line width=1pt] (3,3) circle (1.5cm);

    \node[draw, circle, inner sep=1pt, line width=1pt] at (3,3) {$\times$};

    \draw[thick, decorate, decoration={
        snake,
        amplitude=0.1cm,
        segment length=4mm
      }]
      (3,3) -- (5.5,5.5);

    \node at (3,5.5) {$\varphi\bigl(z\,e^{2\pi i}\bigr)=g\:\cdot\:\varphi(z)$};
  \end{tikzpicture}
  \caption{$\mathsf{M}_g$ monodromy defect imposing $g$-twisted boundary conditions}
  \label{fig:branchcut}
\end{figure}
\noindent
Monodromy defects  have been studied, for example,  as twist fields in $2d$ theories~\cite{Dixon:1985jw,Dixon:1986qv}, as a line  
defect associated to the $\mathbb Z_2$ symmetry of  the Ising $3d$  model \cite{Billo:2013jda},  and  $O(N)$ flavor symmetry  monodromy defects in   free scalars and fermions (see e.g.~\cite{Lauria:2020emq,Giombi:2021uae,Bianchi:2021snj,Bashmakov:2024suh}).

Because the charge conjugation monodromy defect imposes twisted boundary conditions that are detectable at long distances, it is natural to expect that such defects cannot be screened and thus flow to nontrivial conformal defects in the infrared. Thus, charge conjugation defects can be used to define nontrivial conformal defects in a plethora of theories admitting charge conjugation symmetry.

A $\mathbb Z_2^{\mathsf C}$  charge conjugation monodromy defect in ${\cal N}=4$ SYM with Lie algebra $\mathfrak{g}$, which we denote by $\mathsf{M}_{\sigma}$ 
imposes the following twisted boundary conditions:
  \beq
 \Upsilon(z\, e^{2\pi  i})= \sigma[\Upsilon](z)\,,
 \eeq
 where $\sigma[\Upsilon]$ is   an involutive outer autormorphism of $\mathfrak{g}$ and $
\Upsilon=(A_\mu, \lambda, \phi^I)$ are the fields of ${\cal N}=4$ SYM, valued in  $\mathfrak{g}$.\footnote{The quantization of a supersymmetric monodromy defect
for a global symmetry requires a choice of supersymmetric
boundary conditions as the defect is approached. This
  determines which modes of the various fields in
the theory fluctuate in the presence of the monodromy defect.
These boundary conditions need not preserve the full
R-symmetry of the   theory (see e.g. the flavor symmetry
monodromy defect of a hypermultiplet in \cite{Dedushenko:2019yiw}). Our
identification of the $SO(2N)$ monodromy defect with a D3-brane  wrapping $\mathrm{AdS}_3 \times \mathbb{RP}^1\subset \mathrm{AdS}_5 \times \mathbb{RP}^5$
     predicts the
  R-symmetry preserved by the charge conjugation monodromy defect.
} An outer automorphism induces a decomposition
\beq
\mathfrak g={\mathfrak g}^{\sigma}\oplus \mathfrak p\,,
\eeq
where  the subspaces ${\mathfrak g}^{\sigma}$ and $\mathfrak p$ are   the $1$ and $-1$ eigenspaces of the involution $\sigma$ respectively.  Therefore, around such a monodromy defect the fields of   ${\cal N}=4$ SYM obey the following boundary conditions:
\beq
\begin{aligned}
&\Upsilon\in {\mathfrak g}^\sigma:~~~~~~~~~~\text{periodic boundary conditions}\\[+2pt]
 &\Upsilon\in {\mathfrak p}\simeq {\mathfrak g}/{\mathfrak g}^\sigma:~\text{antiperiodic boundary conditions}\,.
\end{aligned}
\eeq

Charge conjugation monodromy defects are labeled\footnote{Of course, we can always stack   a Gukov-Witten surface defect  \cite{Gukov:2006jk} on top of a monodromy defect.} by involutive outer autormorphism of $\mathfrak{g}$ modulo conjugation by  automorphisms of ${\mathfrak g}$. These have been classified by Cartan and are in one-to-one correspondence with (outer) symmetric spaces~\cite{Cartan1926} (see also e.g.~\cite{Helgason2001}\cite{Arias-Tamargo:2019jyh}).\footnote{Outer symmetric spaces  have $\text{rank}({{\mathfrak g}^\sigma)}<\text{rank}({{\mathfrak g})}$ while $\text{rank}({{\mathfrak g}^\sigma)}=\text{rank}({{\mathfrak g})}$ for inner symmetric spaces.} Concretely, for gauge groups that have a holographic dual and charge conjugation symmetry,  we have the following representatives 
\beq
\begin{aligned}
 &\mathfrak{g}=\mathfrak{su}(N) : \qquad \Upsilon(z\, e^{2\pi  i})=\begin{cases}
 -\Upsilon^T(z)\qquad\qquad  \mathfrak{g}^\sigma=\mathfrak{so}(N)~~~~~~~~\qquad\qquad\forall N~~    \\[2pt]
  - J \Upsilon^T(z)J^{-1}\qquad   \mathfrak{g}^\sigma=\mathfrak{sp}(N/2)  ~~~ ~\qquad\qquad N~\text{even}~~
   \end{cases}
  \\[+8pt]
 &\mathfrak{g}=\mathfrak{so}(2N) : \qquad \Upsilon(z\, e^{2\pi  i})= {\cal P}\,\Upsilon(z)\,{\cal P}^{-1} ~~~~\qquad\mathfrak{g}^\sigma=\mathfrak{so}(p)\oplus\mathfrak{so}(2N-p)\qquad p~\text{odd}\,.
 \end{aligned}
 \label{fixedloci}
 \eeq
 where $J$ is the canonical antisymmetric matrix  and  
 \beq
 {\cal P}=\operatorname{diag}\big(\underbrace{-1,\ldots,-1}_{p\ \text{entries}},\underbrace{1,\ldots,1}_{2N-p\ \text{entries}}\big)\,.
\eeq
Since we are considering outer automorphism monodromy defects, $p$ must be odd so that   $\text{det} \, {\cal P}=-1$.  We note that for $\mathfrak{g}=\mathfrak{su}(2N)$  and $\mathfrak{g}=\mathfrak{so}(2N)$ there are  
 distinct monodromy defects  labeled by the choice of  involutive outer autormorphism.

Consider  the   correlation function   of a charge conjugation monodromy defect $\mathsf{M}_{\sigma}$ with a local operator
\beq
\langle \mathsf{M}_{\sigma} \, {\cal O}(x)\rangle\,.
\eeq
Locality  implies that this correlator is non-vanishing only for operators ${\cal O}$ that are invariant under charge conjugation. Consider a charge conjugation invariant operator, the Konishi operator for concreteness.   In the absence of  $\mathsf{M}_{\sigma}$, the one-point function of the Konishi operator vanishes. The insertion of $\mathsf{M}_{\sigma}$ changes  the boundary condition  of the fields in ${\mathfrak g}/{\mathfrak g}^\sigma$ to be antiperiodic while leaving the ones in ${\mathfrak g}^\sigma$ unchanged. The propagator for fields valued in ${\mathfrak g}/{\mathfrak g}^\sigma$ is modified,  resulting in a  non-vanishing one-point function. To leading order:
\beq
\langle \mathsf{M}_{\sigma} \, \hbox{Tr}\left(\phi^I\phi^I\right) \rangle\propto
{\text{dim}\,\mathfrak{g}-\text{dim}\,\mathfrak{g^\sigma}\over |\vec x_{\perp}|^2}\,  \,,
\label{Konishi}
\eeq
where $\mathfrak{g^\sigma}$ is the Lie algebra fixed by the charge conjugation involution $\sigma$. 

It would be interesting to compute the correlation functions of $\mathsf{M}_{\sigma}$ with local operators and 
Wilson and 't Hooft  operators that link $\mathsf{M}_{\sigma}$ using perturbation theory. The correlators with local operators may be amenable to computation via integrability or localization, along the lines of e.g. \cite{Holguin:2025bfe,Chalabi:2025nbg,deLeeuw:2024qki,Choi:2024ktc}. A selection rule can be derived for the correlators with   Wilson and 't Hooft loop operators
\beq
\langle \mathsf{M}_{\sigma}\, W_R \rangle \qquad \langle \mathsf{M}_{\sigma}\,  T_R \rangle\,.
\eeq
Since the linked  loop  operators cross the topological domain wall attached to $\mathsf{M}_{\sigma}$  implementing charge conjugation, which acts by sending $R\rightarrow \mathsf C(R)$, these correlators vanish identically unless $R$ is a self-conjugate representation, that is $R=\mathsf C(R)$.

  \subsection{Holographic Description of Surface Operators $\mathsf{M}_{\mathsf{\sigma}}$}
  \label{sec:holod}
 
We initiate the study  of  the bulk holographic description of  the  charge  conjugation monodromy surface defects $\mathsf{M}_{\sigma}$.  With this goal, it is   insightful to calculate the large $N$ scaling  
of the correlator of $\mathsf{M}_{\sigma}$ with the Konishi operator computed in \eqref{Konishi}.  The correlator scales in the large $N$ limit as\footnote{In the large $N$ limit with $p$ fixed. If $p$ is taken to scale with  $N$, then the correlator scales as $N^2$, like in the $\mathfrak{su}(N)$ theory.}
\beq
 \langle \mathsf{M}_{\sigma} \, \hbox{Tr}\left(\phi^I\phi^I \right) \rangle|\vec x_{\perp}|^2  \propto
 {{\text{dim}\, \mathfrak{g}-\text{dim}\, \mathfrak{g^\sigma}}} \simeq
\begin{cases}
  N^2 \qquad~~ \mathfrak{g}=\mathfrak{su}(N)\\[+4pt]
  N \qquad~~~  \mathfrak{g}=\mathfrak{so}(2N)
\end{cases}\,,
\eeq
where we have used \eqref{fixedloci}. The   correlator \eqref{Konishi} scales   as $N^2$ in the $\mathfrak{su}(N)$ theory, whereas it scales as $N$ in the $\mathfrak{so}(2N)$ theory. This distinction has   important implications for the holographic description of $\mathsf{M}_{\sigma}$.

Our field theory analysis makes the following predictions for the dual description:
\beq
 \mathsf{M}_{\sigma}~\text{in}~\mathfrak{so}(2N)\longleftrightarrow \text{D-brane}
\eeq
and 
\beq
 \mathsf{M}_{\sigma}~\text{in}~\mathfrak{su}(N)\longleftrightarrow \text{geometry}\,.
\eeq
In other words,  $\mathsf{M}_{\sigma}$ in the $\mathfrak{so}(2N)$ theory can be described by a probe D-brane in
the dual background, known to be the $\mathrm{AdS}_5 \times \mathbb{RP}^5$ background~\cite{Witten:1998xy}. On the other hand, in the $\mathfrak{su}(N)$ theory,
the insertion of $\mathsf{M}_{\sigma}$ cannot be described by a D-brane, and should   instead be realized by a new gravitational background.  In arriving at this conclusion, we have used the scaling  of the tension of a D-brane and of Newton's constant  $G_{\textrm{Newton}}$ with $N$ in the 't Hooft limit:
\beq
T_{\text{D-brane}}\sim {1\over g_s}\sim N\,, \qquad \qquad {1\over G_{\textrm{Newton}}}\sim N^2\,.
\eeq

We can be  more precise about the dual description of $\mathsf{M}_{\sigma}$ in the $\mathfrak{so}(2N)$ theory:
\beq
 \mathsf{M}_{\sigma}~\text{in}~\mathfrak{so}(2N)\longleftrightarrow \text{D3-brane(s) on}~\mathrm{AdS}_3 \times \mathbb{RP}^1\subset \mathrm{AdS}_5 \times \mathbb{RP}^5\,.
\eeq
The $\text{D3-brane}$ ends on the boundary of $\mathrm{AdS}_5 \times \mathbb{RP}^5
$ on a maximally symmetric surface, either $\mathbb R^2$ or $S^2$, where $\mathsf{M}_{\sigma}$ is supported. 
This D-brane  was considered previously in the  closely related context of membranes in $\mathrm{AdS}_5$ implementing monodromy   in  \cite{Gukov:1998kn} (see also \cite{GarciaEtxebarria:2022vzq,Etheredge:2023ler}).\footnote{The bulk description of  symmetry defects associated to discrete  global symmetries  of orbifolds of AdS/CFT was considered in \cite{Gukov:1998kn}. The topological defect associated with a charge conjugation transformation was discussed in \cite{Dierigl:2023jdp}. Note that the seven-branes ($\text{4D7+O7}^-$) in \cite{Dierigl:2023jdp}, the $D_4$ singularity
\cite{Sen:1996vd,Vafa:1996xn}, implement  monodromy for the transformation $\Omega (-1)^{F_L}$.  The  $\text{4D7+O7}^-$brane system in 
$\mathrm{AdS}_5 \times S^5$ has been previously considered in \cite{Harvey:2008zz} (see also \cite{Buchbinder:2007ar})}

We now provide   quantitative evidence for this proposal.  The expectation value of any spherical surface defect   is controlled by a surface conformal anomaly $b$~\cite{Graham:1999pm,Schwimmer:2008yh}. In particular, for an spherical monodromy defect of radius $a$, we have~\cite{Drukker:2008wr,Jensen:2015swa,Jensen:2018rxu,Wang:2020xkc,Choi:2024ktc}
\beq
\langle \mathsf{M}_{\sigma}\rangle_{S^2}\propto \left( {a\over a_0} \right)^{b/3}\,,
\eeq
where $a_0$ is a scheme dependent scale.  $b$ is scheme independent.

The expectation is that the contribution to the
$b$ anomaly comes from fields taking values in the
part of the gauge algebra whose propagators are
modified by the surface defect. The value of $b$ is
therefore controlled by the unbroken gauge group,
denoted by $H$, since propagators for fields taking
values in $H$ remain unchanged from those in the
absence of the defect. For a Gukov--Witten surface
defect, $H=\mathbb L$, while for a charge conjugation
monodromy defect, $H=\mathfrak{g}^\sigma$. This leads to
the following value of $b$ for $\mathsf{M}_\sigma$:
\beq
b=3\left[
\dim(\mathfrak g)-\dim(\mathfrak{g}^\sigma)
\right]
\,.
\eeq
 
For the monodromy defect  $\mathsf{M}_{\sigma}$ in the $\mathfrak{so}(2N)$ theory labeled by an integer $p$ (see \eqref{fixedloci}), we have that
\beq
b=3\left[ \dim(\mathfrak{so}(2N))-\dim(\mathfrak{so}(p))- \dim(\mathfrak{so}(2N-p))\right]
\label{formulab}
\eeq
which scales in the large $N$ limit as
\beq
b\simeq 6Np\,.
\label{planaranom}
\eeq

The proposed dual description of  $\mathsf{M}_{\sigma}$ is $p$ D3-branes on $\mathrm{AdS}_3 \times \mathbb{RP}^1$. The corresponding on-shell   action is\footnote{For a closely related computation in the context of Gukov-Witten surface defects, see \cite{Jiang:2024wzs}.}
\beq
S_{\textrm{on-shell}}=p\times T_{D3}\,  L^4\, \text{vol} (\mathrm{AdS}_3 \times \mathbb{RP}^1)
\eeq
where in the $\mathrm{AdS}_5 \times \mathbb{RP}^5$ background\footnote{The exact relation is $L^4=8\pi g_sl_s^4\left(N-{1\over 4}\right)$ and $T_{D3}={1\over (2\pi)^3 g_s l_s^4}$.}
\beq
T_{D3}\,  L^4={N\over \pi^2}
\eeq
to leading order in $N$. Using that the renormalized volume of global $\mathrm{AdS}_3$
is
\beq
\text{vol}(\mathrm{AdS}_3)_\text{ren}=2\pi \log(a_0/a)
\eeq
we find that
\beq
S_{\textrm{on-shell}}=2pN \log(a_0/a)\,.
\eeq
Translating into field theory,  we find 
\beq
\langle \mathsf{M}_{\sigma}\rangle_{S^2}=e^{-S_{\textrm{on-shell}}}=\left( {a\over a_0} \right)^{2pN}\,,
\eeq
which  exactly reproduces the planar   monodromy surface defect conformal anomaly \eqref{planaranom}.

We now turn to the dual gravitational description of  $\mathsf{M}_{\sigma}$  in the $\mathfrak{su}(N)$ theory. The large $N$ behaviour of correlators of $\mathsf{M}_{\sigma}$ with local operators in the $\mathfrak{su}(N)$ theory predicts that the insertion of $\mathsf{M}_{\sigma}$ is so ``heavy"  that it must described by a new string background.
We expect that the dual geometries will be closely related to the bubbling geometries for Gukov-Witten surface operators~\cite{Gomis:2007fi,Lin:2005nh}.\footnote{For the gravitational description of supersymmetric monodromy defects in ${\cal N}=4$ SYM associated to continous global symmetries see \cite{Arav:2024exg}.  See also e.g. \cite{Bomans:2024vii}.}

Our field theory analysis   provides further important clues about the dual gravitational description of $\mathsf{M}_{\sigma}$. As we described, there are two distinct monodromy defects in the  
$\mathfrak{su}(N)$ theory, depending on the choice of which $\mathfrak{g}^\sigma$ is fixed by the  involutive 
outer automorphism. The one-one loop computation in \eqref{Konishi} yields the following functions of N
\beq
\begin{aligned}
&\mathfrak{g}^\sigma= \mathfrak{so}(N)\rightarrow \text{dim}\,\mathfrak{g}-\text{dim}\,\mathfrak{g}^\sigma= \frac{N^2 + N - 2}{2}\\
 &\mathfrak{g}^\sigma= \mathfrak{sp}(N/2)\rightarrow \text{dim}\,\mathfrak{g}-\text{dim}\,\mathfrak{g}^\sigma=\frac{N^2 - N - 2}{2}\,.
\end{aligned}
\eeq
There are two noteworthy observations that follow from this:
\begin{enumerate}
\item even though the familiar  correlation functions in the $\mathfrak{su}(N)$ theory have a $1/N^2$ expansion organized as a sum over oriented Riemann surfaces,  correlators of     $\mathsf{M}_{\sigma}$ have a $1/N$ expansion and include sums over unoriented Riemann surfaces, including the leading unoriented surface $\mathbb{RP}^2$
\item the correlators of  the two distinct $\mathsf{M}_{\sigma}$ in the $\mathfrak{su}(N)$ theory are mapped into each other by the transformation $N\rightarrow -N$
\end{enumerate}

These  are a consequence of the fact that the propagators of the $\mathfrak{su}(N)$ theory in the presence of $\mathsf{M}_{\sigma}$ are ``twisted" in the double line notation, which result in a large $N$ expansion organized by unoriented Riemann surfaces.  And the   way the propagators are twisted  depends on the choice of 
$\mathsf{M}_{\sigma}$, and parallels the well-known relation between the large $N$ expansion of $\mathfrak{so}(N)$ and $ \mathfrak{sp}(N/2)$ theories~\cite{Cicuta1982TopologicalExpansionSOSp}. 

This makes a   sharp prediction for the gravitational duals of the   monodromy defects $\mathsf{M}_{\sigma}$ in the $\mathfrak{su}(N)$ theory (cf.~\cite{Ambrosino:2026ovo}). The dual string theory  must involve unoriented strings, that is the dual geometric background must involve  worldsheet parity $\Omega$. The crosscap state corresponding to the orientifold projection captures the contribution from unoriented Riemann surfaces. The two distinct choices of $\mathsf{M}_{\sigma}$, which are mapped  to each other by $N\rightarrow -N$, are realized in the dual unoriented string  theory background as the two possible choices of sign of the crosscap state. The string perturbative expansion for the two choices differ by a sign precisely for the unoriented Riemann surfaces that have an odd number of crosscaps. Our candidate closed string background dual to $\mathsf{M}_{\sigma}$ is the backreaction of $N$ ${\rm D3}$-branes in $\mathbb R^{1,3}$ with an ${\rm O3}^\pm$-plane that is codimension-two in $\mathbb R^{1,3}$, which indeed defines a codimension-two defect in ${\cal N}=4$ SYM (this brane configuration has  appeared in  \cite{Gukov:2008sn,Arias-Tamargo:2022nlf}). 
It would be interesting to further develop this identification.

  \section{Discussion}
\label{sec:summary}

In this paper, we have studied the realization of the fundamental discrete symmetries in quantum field theory -- namely, $\mathsf{C}$, $\mathsf{P}$, and $\mathsf{T}$ --  within ${\cal N}=4$ SYM. We have shown that 
 $\mathsf{P}$  and $\mathsf{T}$ symmetry transformations can be defined but necessarily do not  commute with the $SU(4)$ R-symmetry, unlike $\mathsf{C}$.
By analyzing how these discrete symmetries act on ${\cal N}=4$ SYM defect operators, we have found that under S-duality:
\[
\begin{aligned}
  \mathsf{P} &\longleftrightarrow \mathsf{CP}, \\
  \mathsf{T} &\longleftrightarrow \mathsf{CT},
\end{aligned}
\]
where $S$ must be accompanied by an outer automorphism of the $SU(4)$ R-symmetry   \eqref{outerSU4}.
$\mathsf P$ and $\mathsf T$ in one duality frame become  $\mathsf{CP}$ and $\mathsf{CT}$ in the S-dual frame,
while $\mathsf C$ and $ \mathsf{CPT}$ are invariant under S-duality.\footnote{Dualities   
exchanging  orientation reversing symmetries include $3d$ infrared dualities \cite{Seiberg:2016gmd} and $S$-duality in electromagnetism \cite{Metlitski:2015yqa}.}  The strategy of defining $\mathsf{P}$ and $\mathsf{T}$ symmetries by combining spacetime transformations with inversions in the R-symmetry,  can  be extended to other gauge theories arising from the dimensional reduction of chiral gauge theories. One can also deduce, in a similar fashion we did, the        transformation of these symmetries under dualities. This includes, in particular,  certain $4d$ ${\cal N}=2$ theories that originate as the dimensional reduction of $6d$ ${\cal N}=1$ theories. Since $4d$   theories that are spin,\footnote{Spin theories are theories that require the manifold to admit an spin structure to be defined.} have no 't Hooft anomalies for orientation-reversal symmetries, due to the vanishing of the cobordism groups $\Omega^{\mathrm{Pin}^\pm}_5$,     these symmetries  can be consistently gauged to define the theory on unoriented $\mathrm{Pin}^\pm$ manifolds.
 The existence of a gamut of orientation reversing symmetries implies that ${\cal N}=4$ SYM can be studied on unoriented manifolds with a   $\mathrm{Pin}^+$ and $\mathrm{Pin}^-$ structure. 
Studies of $4d$ gauge theories on $\mathbb{R}P^4$, which is a  $\mathrm{Pin}^+$ manifold, include \cite{Metlitski:2015yqa,LeFloch:2017lbt,Wang:2020jgh,Caetano:2022mus,Zhou:2024ekb}.  It would be interesting to study ${\cal N}=4$ SYM (and other gauge theories having a $\mathsf{T}^2=1$ symmetry) on $\mathrm{Pin}^-$ manifolds, such as $S^2\times  \mathbb{R}P^2$.
   
We have established a fundamental entry in the dictionary of the AdS/CFT correspondence,   identifying the symmetries of Type IIB string theory in $AdS_5\times S^5$ that are dual to $\mathsf{C}$, $\mathsf{P}$, and $\mathsf{T}$ in $\mathfrak{su}(N)$ ${\cal N}=4$ SYM:
\beq
\begin{aligned}
  \mathsf{C} &\leftrightarrow \Omega (-1)^{F_L} R_{456789}, \\[2pt]
  \mathsf{P} &\leftrightarrow (-1)^{F_L} R_{3456}, \\[2pt]
  \mathsf{T} &\leftrightarrow \Omega R_{0789}\,.
 \end{aligned}
\label{CPTfin}
\eeq
The mapping involves ``stringy"    symmetries  as well as geometric transformations.
  $\Omega$ denotes worldsheet parity, $(-1)^{F_L}$  is spacetime fermion parity carried by left-movers on the string worldsheet, and $R_{M}$ is   reflection  along the $x^M$ coordinate of $AdS_5\times S^5$ (cf. \eqref{metricads}).
We  determined \eqref{CPTfin}   using  the bulk description of defect operators in ${\cal N}=4$ SYM in terms of supergravity modes, branes in $AdS_5\times S^5$ and asymptotically $AdS_5\times S^5$ bubbling solutions of Type IIB string theory. The proposed   symmetries act on the bulk description of ${\cal N}=4$ SYM defect operators precisely as 
$\mathsf{C}$, $\mathsf{P}$, and $\mathsf{T}$ does.

While     the $\mathsf{CPT}$ theorem guarantees  $\mathsf{CPT}$  symmetry    in any  relativistic quantum field theory, the situation is not as clear in quantum gravity.\footnote{See \cite{McNamara:2022lrw,Harlow:2023hjb,Witten:2025ayw} for recent discussions of $\mathsf{C}$-$\mathsf{P}$-$\mathsf{T}$ symmetries in quantum gravity.} There is no equally sharp gravitational counterpart of the   $\mathsf{CPT}$ theorem, but it is generally expected that some version  of the $\mathsf{CPT}$ theorem holds in quantum gravity, at least for certain asymptopia. In perturbative string theory vacua with suitable asymptotic backgrounds, a  target space  $\mathsf{CPT}$ theorem can be derived from the string worldsheet by implementing -- on the worldsheet fields -- the   Wick rotation used to derive the quantum field theory  $\mathsf{CPT}$ theorem (see e.g. \cite{Polchinski:1998rr}). In the context of holography, at least in asymptotically $AdS$ backgrounds, 
$\mathsf{CPT}$  symmetry is expected to be a symmetry of the bulk quantum gravity theory. In this paper we have seen, that $\mathsf{CPT}$  symmetry  of the boundary theory {\it is} $\mathsf{CPT}$  symmetry in bulk quantum gravity:
 \beq
 \mathsf{CPT}\longleftrightarrow     \mathsf{CPT}_{\textrm{bulk}}\,.
 \label{canoCPT}
 \eeq
It is natural to expect that, while the bulk realization of $\mathsf{C}$-$\mathsf{P}$-$\mathsf{T}$ symmetries  in  each holographic duality is definitely model dependent, that $\mathsf{CPT}$ symmetries of dual  bulk and boundary theories are canonically identified as in \eqref{canoCPT}. It would be interesting to map out  $\mathsf{C}$-$\mathsf{P}$-$\mathsf{T}$ symmetries across other holographic dual pairs. 

Do the transformations dual to $\mathsf{C}$, $\mathsf{P}$, and $\mathsf{T}$ generate global symmetries or ``gauge symmetries" in the bulk? It is widely expected that all true symmetries in quantum gravity  are gauged~(see e.g. \cite{Banks:2010zn,Witten:2017hdv,Harlow:2018tng}). The expectation that quantum gravity has no global symmetries has  been  formalized in the Cobordism Conjecture~\cite{McNamara:2019rup}. This expectation is   robust for continuous symmetries and relies on rather well-understood physics of black holes.  Consider throwing a charged particle into a black hole. If the charge is associated with a gauge symmetry, then despite the particle being hidden behind the horizon, the ``Faraday lines" emanating from it extend outside the horizon, allowing the gauge charge to be measured via a surface integral   outside the black hole.\footnote{On-shell, the Noether current associated with a gauge transformation is a total derivative. Consequently, by Stokes' theorem, the corresponding gauge charge can be measured by an integral over a codimension-two surface. This contrasts with a global charge, which is measured by integrating the associated Noether current over a codimension-one surface.
} In contrast, if the charge is associated with a global symmetry, it is lost once it crosses the horizon, since no external field lines are associated with global charges. Thus, global charge conservation is violated, supporting the expectation that exact global symmetries do not exist in quantum gravity.

Unlike continuous gauge symmetries, discrete gauge symmetries do not have associated massless gauge fields, which makes the black hole argument outlined above less airtight.
 Even in a non-gravitational setting, it is subtle to distinguish {\it locally} a discrete global symmetry from a discrete gauge symmetry.  Operationally, on a manifold with nontrivial topology, one quantizes the theory in the presence of a fixed background gauge field for a global symmetry. In contrast, if the symmetry is gauged, one must sum over all gauge field configurations. To a gauged discrete symmetry, one can generically associate a codimension-two defect
 around which fields are subjected to a gauge transformation.

Irrespective  of black hole arguments,   the expectation in AdS/CFT is that all global symmetries of the boundary CFT, continuous or discrete, correspond to asymptotic gauge symmetries in the bulk gravitational theory~\cite{Witten:1998qj}. 
In the AdS/CFT correspondence, given a spacetime $X$ on which the CFT is defined, we are instructed to sum -- at least in the semiclassical bulk approximation --  over all bulk geometries $Y$ which asymptote to $X$.  In the presence of a fixed background gauge field for a global symmetry  $H$ of the boundary CFT on $X$, the duality requires extending the background gauge field over $Y$. But, just as we  must sum over all spacetimes $Y$ with prescribed asymptopia $X$, we must also sum over all extensions of the connection over $Y$ with prescribed  Dirichlet  boundary conditions on $X$. This sum over extensions implies that the background gauge field becomes dynamical in the bulk, defining a gauge theory with gauge group $H$ in the gravitational theory. In other words, the bulk theory contains a fluctuating gauge field with gauge group $H$, indicating that the symmetry is gauged in the bulk. For other arguments using  AdS/CFT that  all bulk symmetries are gauged,   see \cite{Harlow:2018tng}.

One can argue directly that the bulk transformations   appearing in the AdS/$\mathsf{C}$-$\mathsf{P}$-${\mathsf T}$~Correspondence \eqref{CPTfin} are gauged. Indeed, $R_M$, which implements an involution in the target space geometry, is part of the target space (asymptotic) diffeomorphism group acting on $AdS_5\times S^5$. As such, it is a target space gauge transformation. How about the worldsheet symmetries $\Omega$ and $(-1)^{F_L}$? A convincing argument that they correspond to  gauge transformations of the target space theory, uses  the duality between Type IIB  string theory with M-theory on $T^2$ \cite{Schwarz:1995dk,Schwarz:1995jq} (or the duality with F-theory~\cite{Vafa:1996xn}). 
$\Omega$ and $(-1)^{F_L}$ then  correspond to a parity transformation on $T^2$~\cite{Dabholkar:1996pc}, and once again, it is an spacetime diffeomorphism, and the symmetries are gauged in the target space. 

In the boundary theory, assuming there are no 't Hooft anomalies for $H$, one can gauge the global symmetry $H$. Since this symmetry is already gauged in the bulk, a natural question arises: what is the bulk counterpart of gauging a global symmetry in the boundary theory? The answer is that gauging $H$ on the boundary corresponds to modifying the boundary condition of the bulk gauge field associated with $H$ from a Dirichlet to a Neumann boundary condition \cite{Witten:1998wy,Aharony:2016kai}.\footnote{Examples of how changing boundary conditions   gauges boundary symmetries  include \cite{Witten:2003ya,Bergman:2022otk}.}
 In other words, instead of fixing the value of the gauge field at the boundary, one allows it to fluctuate, making it dynamical in the boundary, thus gauging the boundary symmetry.

Consider ${\cal N}=4$ SYM with gauge group $SU(N)$. Since the spin-cobordism group $\Omega_5^{\text{Spin}}(B\mathbb Z_2)$ is trivial, the $\mathbb Z^{\mathsf C}_2$ charge conjugation symmetry of any  $4d$ quantum field theory that is spin can be gauged.\footnote{Since $\Omega_5^{\text{SO}}(B\mathbb Z_2)$ is nontrivial, gauge theories that are bosonic, in the sense that they do not require a choice of spin structure, could have an obstruction to gauging $\mathbb Z^{\mathsf C}_2$.} Gauging charge conjugation $\mathsf C$  changes the gauge group from $SU(N)$ to the disconnected group $SU(N)\rtimes \mathbb Z_2^{\mathsf C}$, which fits in the long exact sequence:\footnote{For  recent work on $4d$ gauge theories with a disconnected gauge group, see e.g. \cite{Bourget:2018ond,Arias-Tamargo:2019jyh}.}
\beq
0\rightarrow SU(N)\rightarrow SU(N)\rtimes \mathbb Z_2^{\mathsf C}\rightarrow \mathbb Z_2^{\mathsf C}\rightarrow 0\,.
\eeq 
This corresponds to taking the bulk $\mathbb Z_2^{\mathsf C}$ gauge field in $AdS_5$ to obey Neumann boundary conditions (cf. \cite{Aharony:2016kai}).

The    question formulated in the   paragraph above is interesting and somewhat confusing  in   perturbative string theory. First, it is believed that all discrete global symmetries of the theory on the string worldsheet uplift to symmetries of the target space theory. It is usually stated \cite{Banks:1988yz,Banks:2010zn,Polchinski:1998rr} that global symmetries on the string worldsheet correspond to gauge symmetries of the target space theory, ultimately some version of string field theory.  
But if the   worldsheet theory has a discrete global symmetry $H$, and this symmetry is gauged in the target space theory, what {\it operation} in the target theory corresponds 
to gauging $H$ on the worldsheet? It would be desirable to have a general answer to this question.
Consider the following  illustrative example. The string worldsheet     theory on 
$AdS_5\times  S^5$ has an $\Omega (-1)^{F_L} R_{456789}$ global symmetry. This is also a symmetry of the target space Type IIB string field theory on $AdS_5\times  S^5$.  As discussed above,  gauging ${\mathsf C}$ in ${\cal N}=4$ SYM changes the gauge group to $SU(N)\rtimes \mathbb{Z}_2^{\mathsf C}$, and corresponds to modifying the boundary conditions in the target space theory for the $\mathbb{Z}_2$ gauge field associated with the $\Omega (-1)^{F_L} R_{456789}$ gauge transformation.
 On the other hand, gauging $\Omega (-1)^{F_L} R_{456789}$ on the string worldsheet yields Type IIB string theory on  $AdS_5\times  \mathbb{R}P^5$ \cite{Witten:1998xy}. This is dual, instead,  to ${\cal N}=4$ SYM with Lie algebra $\mathfrak{so}/\mathfrak{sp}$ \cite{Witten:1998xy}. This highlights the important difference between gauging a symmetry on the string worldsheet and a symmetry being a gauge symmetry in the target space string field theory.

We have also  initiated  the study of codimension-two charge conjugation monodromy defects  
in quantum field theory. We have defined       surface defects   in ${\cal N}=4$ SYM and deduced some of its properties and analyzed its simplest correlators. We have   advanced aspects of the bulk gravitational description of  these defects and subjected the proposal to various checks. 
 It would be desirable to use 
  localization, integrability and the bootstrap to compute, perhaps exactly, correlators of monodromy defects  with bulk ${\cal N}=4$ SYM operators and surface defect operators.\footnote{For localization 
computations of surface defects on $S^4$,
see e.g.~\cite{
Gomis:2014eya,
LamyPoirier:2014Surface,
GaiottoKim:2014Surface,
GomisEtAl:2016Intersecting,
PanPeelaers:2016Intersecting,
PanPeelaers:2017Chiral,Choi:2024ktc},  and for integrability
in the presence of surface defects, see
e.g.~\cite{deLeeuw:2024qki,Holguin:2025bfe,Chalabi:2025nbg}.}

Since charge conjugation monodromy   defect operator implements twisted boundary conditions that can be detected from
afar,  one can expect that charge conjugation defects cannot be screened, and flow in the infrared to conformal defects. 
Since charge conjugation symmetry is rather ubiquitous in all dimensions,  charge conjugation monodromy defects
provide an ultraviolet definition of putative conformal infrared defects. Also,   charge  conjugation symmetry is
realized differently across infrared dualities, and therefore, the infrared conformal defect admits multiple short distance realizations. It would be interesting to explore charge conjugation monodromy defects in various field theories of interest using state of the art methods and elucidate their role and across   infrared dualities. It would also
be desirable to understand the construction of monodromy defects associated to spacetime symmetries such as $\mathsf{P}$ and $\mathsf{T}$.

 \medskip\medskip
\section*{Acknowledgments}

The author  would like to thank C. Choi, D. Gaiotto, J. McNamara, A. Sen and Y. Wang  for   discussions.
We would like to thank the organizers of ``Program on Anomalies, Topology and Quantum Information in Field Theory and Condensed Matter Physics", where the results of this paper were  presented. 
Research at Perimeter Institute is supported in part by the Government of Canada through the Department of Innovation, Science and Economic Development Canada and by the Province of Ontario through the Ministry of Colleges and Universities.

 \vfill\eject
 
\end{fmffile}
\bibliography{refs}
\end{document}